\RequirePackage{fix-cm}
\documentclass[twocolumn]{svjour3}          
\smartqed  
\usepackage{graphicx}
\usepackage[utf8]{inputenc}
\usepackage{latexsym}
\usepackage{amssymb}
\usepackage{amsmath}
\usepackage{graphics}
\usepackage{float}
\usepackage{mcite}

\journalname{EPJ C}

\begin{document}
	
	\title{Odderon effects in the differential cross-sections at Tevatron and LHC energies
	}
	
	
	\author{Evgenij Martynov         \and
		Basarab Nicolescu 
	}
	
	
	\institute{Evgenij Martynov \at
		Bogolyubov Institute for Theoretical Physics, Metrologichna 14b, Kiev, 03680 Ukraine \\
		\email{martynov@bitp.kiev.ua}           
		\and
		Basarab Nicolescu \at
		Faculty of European Studies, Babes-Bolyai University, Emmanuel de Martonne Street 1, 400090 Cluj-Napoca, Romania\\
		\email{basarab.nicolescu@gmail.com}           
	}
	
	\date{Received: date / Accepted: date}

	\maketitle
	
	\begin{abstract}
In the present paper, we extend the Froissa\-ron-Maximal Odderon (FMO) approach at $t$ different from 0. Our extended FMO approach gives an excellent description of the 3266 experimental points considered in a wide range of energies and momentum transferred. We show that the very interesting TOTEM results for proton-proton differential cross-section in the range 2.76-13 TeV, together with the Tevatron data for antiproton-proton at 1.8 and 1.96 TeV give further experimental evidence for the existence of the Odderon. One spectacular theoretical result is the fact that the difference in the dip-bump region between $\bar pp$ and $pp$ differential cross-sections is diminishing with increasing energies and for very high energies (say 100 TeV), the difference between $\bar pp$ and $pp$ in the dip-bump region is changing its sign: $pp$ becomes bigger than $\bar pp$ at $|t|$ about 1 GeV$^2$. This is a typical Odderon effect. Another important - phenomenological - result of our approach is that the slope in $pp$ scattering has a different behavior in $t$ than the slope in $\bar pp$ scattering. This is also a clear Odderon effect.	
\end{abstract}

\section{Introduction}
\label{intro}
The Odderon is certainly one the most important problems in strong interaction physics. It was introduced \cite{LN} in 1973 on the basis of asymptotic theorems \cite{AsymptTheorems}, \cite{Cornille} and was rediscovered later in QCD \cite{Odderon-B, Odderon-KP, OdderonQCD-1, OdderonQCD-2, OdderonQCD-R}. In spite of the fact that its theoretical status is very solid, its experimental evidence from half a century is still scarce. This situation is not astonishing, The clear evidence for Odderon has to come by comparing the data at the same energy  in hadron-hadron and antihadron-hadron scatterings. But we have not such accelerators!  We therefore have to limit our search for evidence for  the Odderon  only in an indirect way. The search for the Odderon is crucial in order to confirm the validity of QCD.
It is very fortunate that the TOTEM datum $\rho^{pp} = 0.1\pm 0.01$ at 13 TeV \cite{TOTEM-01} is the first  experimental  discovery of the Odderon at $t=0$, namely in its maximal form  \cite{MN-0}. Moreover, we checked recently that just the Maximal Odderon in FMO approach is preferred by the experimental data. We generalized the FMO approach by relaxing the $\ln^2s$ constraints both in the even- and odd-under-crossing amplitude and we show that, in spite of a considerable freedom of a large class of amplitudes, the best fits bring us back to the maximality of strong interaction \cite{MN-1}.

In the present paper, we extend the FMO approach at $t$ different from 0.  We show that the very interesting TOTEM results for proton-proton differential cross-section in the range 2.76-13 TeV, together with the D0 data for antiproton-proton at 1.96 TeV give further experimental evidence for the existence of the Odderon.

\section{Extension of the FMO approach at $t$ different from zero - General definitions} 
\label{sec:1}
In general amplitude of $pp$ forward scattering is
\begin{equation}\label{eq:ppampl0}
F_{pp}(s,t) = F_+(s,t) + F_-(s,t) 
\end{equation}
and the amplitude of antiproton-proton scattering is			
\begin{equation}\label{eq:papampl0}
F_{\bar pp}(s,t) = F_+(s,t) - F_-(s,t).
\end{equation}
In this model we used the following normalization of the physical amplitudes. 
\begin{equation}\label{eq: observ}
\begin{array}{ll}
\sigma_t(s)&=\dfrac{1}{\sqrt{s (s-4m^2 )}}\text{Im} F(s,0), \\
\dfrac{d\sigma_{el}}{dt}&=\dfrac{1}{64\pi ks(s-4m^2)}|F(s,t)|^2
\end{array}
\end{equation}
where $k=0.3893797\,\, \text{mb}\cdot\text{GeV}^2$. With this normalization the amplitudes have dimension $\text{mb}\cdot\text{GeV}^2$.  

Strictly speaking crossing-even (CE), $F_+(s,t)$, and cros\-sing-odd (CO),  $F_-(s,t)$, parts of amplitudes are defined as functions of $z_t=(t+2s-4m^2)/(4m^2-t)$, where $m$ is proton mass, with the property 
\begin{equation}\label{eq:crossym}
F_{\pm} (-z_t,t)=\pm F_{\pm} (z_t,t). 
\end{equation}
In the FMO model CE and CO terms of amplitudes are defined as sums of the asymptotic contributions $F^H(s,t)$,  $F^{MO}(s,t)$  and Regge pole contributions which are important at the intermediate and relatively low energies
\begin{equation}\label{eq:FMO+R}
\begin{array}{l}
F_+(z_t,t)=F^H(z_t,t)+F^{R_+}(z_t,t), \\
F_-(z_t,t)=F^{MO}(z_t,t)+F^{R_-}(z_t,t)
\end{array}
\end{equation} 
where $F^H(z_t,t)$ denotes the Froissaron contribution and $F^{MO}(z_t,t)$ denotes the Maximal Odderon contribution. Their specified form will be defined below.

\section{Regge poles and their double rescatterings}
\label{sec:2}
In the FMO model in the terms $F^{R_{\pm}}(s,t)$ we consider not only single Regge pole contributions but also  their double rescatterings or double cuts. Their contributions,   $F^R_{pp}(z_t,t), F^R_{\bar pp}(z_t,t)$, are the following
\begin{equation}\label{eq:bare-0}
\begin{array}{ll}
F^R_{pp}(z_t,t)=F^{_+}(z_t,t)+F^{_-}(z_t,t),\\
F^R_{\bar pp}(z_t,t)=F^{_+}(z_t,t)-F^{_-}(z_t,t)\\
\end{array}
\end{equation} 
where  $z_t=-1+2s/(4m^2-t)\approx 2s/(4m^2-t)$. For convenience in further work with parameterizations in FMO model at $t=0$ and $t\neq 0$ contrary to standard definition of $z_t$ we put opposite sign for it.
\begin{equation}\label{eq:bare-1}
\begin{array}{ll}
F^{_+}(z_t,t)&=F^P(z_t,t)+F^{R_+}(z_t,t)+F^{PP}(z_t,t)\\
&+F^{OO}(z_t,t),\\
F^{_-}(z_t,t)&=F^O(z_t,t)+F^{R_-}(z_t,t)+F^{PO}(z_t,t).
\end{array}
\end{equation} 
Here $F^P(z_t,t), F^O(z_t,t)$ are simple $j$-pole Pomeron and Odderon contributions and $F^{R_+}(z_t,t),  F^{R_-}(z_t,t)$ are effective $f$ and $\omega$ simple $j$-pole contributions, where $j$ is an angular momenta of these reggeons. $F^{PP}(z_t,t)$,  $F^{OO}(z_t,t),$ $F^{PO}(z_t,t),$ are double $PP, OO, PO$ cuts, correspondingly. We consider the model at $t\neq 0$ and at energy $\sqrt{s}> 19$ GeV, so we neglect the rescatterings  of secondary reggeons with $P$ and $O$. In the considered kinematical region they are small. Besides, because $f$ and $\omega$ are effective, they  can take into account small effects from the cuts.  The standard Regge pole contributions have the form
\begin{equation}\label{eq:sec Regge}
F^{R_\pm}(z_t,t)=-\binom{1}{i}
2m^2C^{R_\pm}(t)(-iz_t)^{\alpha_\pm(t)}
\end{equation}
where $R_\pm= P,O,R_+,R_-$ and $\alpha_P(0)=\alpha_O(0)=1$. The factor 
$2m^2$ is inserted in amplitudes  $F^{R_\pm}(z_t,t)$ in order to have the normalization for amplitudes and dimension of coupling constants (in mb) coinciding with those in \cite{MN-0}. The same is made for all other amplitudes, including Froissaron and Maximal Odderon (see below).  

For the coupling function $C^{R_\pm}(t)$ we have considered two possibilities. The first one is a simple exponential form. It is used for the secondary reggeons, because we did not consider low energies where terms $R_{\pm}(s,t)$ are more important.  
\begin{equation}\label{eq:nonexpvert}
C^{R_\pm}(t)=C^{R_\pm}e^{b^{R_{\pm}}t}, \qquad C^{R_\pm}(0)=C^{R_\pm}.
\end{equation}
The second case is a linear combination of exponents for Standard Pomeron and Odderon terms  which allow to take into account some possible effects of non-exponential behavior of coupling function. \begin{equation}\label{eq:nonexp-coupling}
\begin{array}{ll}
C^{P,O}(t)&=C^{P,O}\left [\Psi^{P,O}(t) \right ]^2, \\
 \Psi^{P,O}(t)&=d_{p,o}e^{b_1^{P,O}t}+(1-d_{p,o})e^{b_2^{P,O}t}.
\end{array}
\end{equation}

We have added as well the double pomeron and odderon cuts, $PP, OO, PO$ in their exact form  without any new parameters.  Namely,
\begin{equation}\label{eq:PP}
\begin{array}{ll}
F^{PP}(z_t,t)&=-i\dfrac{(z_tC^P)^2}{16\pi s\sqrt{1-4m^2/s}} \left \{ \dfrac{d_p^2}{2B_1^p}\exp(tB_1^p/2)\right .\\ 
&
+\dfrac{2d_p(1-d_p)} {B_1^p+B_2^p}\exp\left (t\dfrac{B_1^pB_2^p}{B_1^p+B_2^p} \right )\\
&\left .+\dfrac{(1-d_p)^2} {2B_2^p}\exp(tB_2^p/2) \right \} 
\end{array}
\end{equation}
\begin{equation}\label{eq:OO}
\begin{array}{ll}
F^{OO}(z_t,t)&=-i\dfrac{(z_tC^O)^2}{16\pi s\sqrt{1-4m^2/s}} 
 \left \{ \dfrac{d_o^2}{2B_1^o}\exp(tB_1^o/2)\right . \\ 
&+\dfrac{2d_o(1-d_o)} {B_1^o+B_2^o}\exp\left (t\dfrac{B_1^oB_2^o}{B_1^o+B_2^o} \right )+\\
&\left .\dfrac{(1-d_o)^2} {2B_2^o}\exp(tB_2^o/2) \right \} 
\end{array}
\end{equation}
where $B_k^{p,o}=b_k^{P.O}+\alpha'_{P,0}\ln(-iz_t),\quad k=1,2 , \quad b_k^{P,O}$ are the constants from single pomeron and odderon contributions.
\begin{equation}\label{eq:PO}
\begin{array}{ll}
F^{PO}(z_t,t)&=\dfrac{z_t^2C^PC^O}{16\pi s\sqrt{1-4m^2/s}} \\
&\times \left \{
\dfrac{d_pd_o}{B_1^p+B_1^o}\exp\left (t\dfrac{B_1^pB_1^o}{B_1^p+B_1^o}\right )\right .\\ 
&+ \dfrac{d_p(1-d_o)}{B_1^p+B_2^o}\exp\left (t\dfrac{B_1^pB_2^o}{B_1^p+B_2^o}\right)\\
& +\dfrac{(1-d_p)d_o}{B_2^p+B_1^o}\exp\left(t\dfrac{B_2^pB_1^o}{B_2^p+B_1^o}\right )+\\
&\left . \dfrac{(1-d_p)(1-d_o)}{B_2^p+B_2^o}\exp\left(t\dfrac{B_2^pB_2^o}{B_2^p+B_2^o}\right) \right \}  
\end{array}
\end{equation}
We have found that for a better description of the data it is reasonable to add to the amplitudes the  contributions which mimic some  properties of ''hard`` pomeron ($P^H$) and  odderon ($O^H$). We take  them in the simplest form
\begin{equation}
P^H(t)=i\dfrac{C^{PH}z_t}{(1-t/t_P)^{\mu_P} }, \quad \mu_P\leq 4.
\end{equation}
\begin{equation}
P^O(t)=\dfrac{C^{OH}z_t}{(1-t/t_O)^{\mu_OP} }, \quad \mu_O\leq 4.
\end{equation}

\section{Froissaron and Maximal Odderon at $t\neq 0$}
\label{sec:3}

\subsection{Partial amplitudes for Froissaron and Odderon}
\label{sec:4}
Let us start from the Froissaron amplitude in  ($s,t$)-repre\-sen\-tation at high $s$. 
The amplitude can be expanded in the series of partial amplitudes $\phi(\omega,t)$. In accordance with the standard definition of partial amplitude
\begin{equation}\label{eq:definition-of-partial-amplitudes}
F(z_t,t)=16\pi \sum\limits_{j=0}^{\infty}(2j+1)P_j(-z_t)\phi(j,t).
\end{equation}
With such definition partial amplitude satisfies the unitarity equation in the form
\begin{equation}\label{eq:unitarity4partial-amplitudes}
\begin{array}{ll}
\text{Im}\phi(j,t)& = \rho(t)|\phi(j,t)|^2+\text{inelastic contribution},\\\\
\rho(t)&=\sqrt{1-4m^2/t}
\end{array}
\end{equation}

We use of the Sommerfeld-Watson transform amplitude (here and in what follows $\omega=j-1$ and $j$ is complex angular momentum) which can be written as follows  
\begin{equation}\label{eq:mellin-represantation}
\begin{array}{ll}
F^\zeta(z_t,t)&=16\pi\sum\limits_{\xi=-1,1} \int\limits_C\dfrac{d\omega}{2\pi i} (2\omega+3)\dfrac{1-\xi e^{-i\pi \omega}}{-\sin(\pi\omega) }\\
&\times \phi^\xi(\omega,t) P_{1+\omega}(z_t) \\  
&=16\pi\sum\limits_{\xi=-1,1} \int\limits_C\dfrac{d\omega}{2\pi i} (2\omega+3)\\
&\times e^{-i\pi \omega/2}\dfrac{e^{i\pi \omega/2}-\xi e^{-i\pi \omega/2}}{-\sin(\pi\omega)}\phi^\xi (\omega,t)P_{1+\omega}(z_t)\\
&=z_t\sum\limits_{\xi=-1,1} \int\limits_C\dfrac{d\omega}{2\pi i}e^{\omega \zeta}\varphi^\xi(\omega,t).
\end{array}
\end{equation}
where $\omega=j-1$, $\xi$ is the signature of the term, contour $C$ is a straight line parallel to imaginary axis and at the right of all singularities of $\phi^\xi(\omega,t)$,  $\zeta=\ln(z_t)-i\pi/2 \equiv \ln(-iz_t)$ and 
\begin{equation}\label{eq:ext-partial-amplitude}
\begin{array}{ll}
\varphi^\xi(\omega,t)&=16\pi (2\omega+3)\dfrac{e^{i\pi \omega/2}-\xi e^{-i\pi \omega/2}}{-\sin(\pi \omega)}\pi^{-1/2}2^{ \omega+1}\\
&\times \dfrac{\Gamma(\omega+3/2)}{\Gamma(\omega+2)}\phi^\xi(\omega,t) 
\end{array}
\end{equation}
Thus for crossing even amplitude ($\xi$=+1) we have
\begin{equation}\label{eq:CE-part-ampl-xi=1}
\varphi^+(\omega,t)=i32\sqrt{\pi}(2\omega+3)\dfrac{\Gamma(\omega+3/2)}{\Gamma(\omega+2)}2^\omega \dfrac{\phi^+(\omega,t)}{\cos(\pi \omega/2)}
\end{equation}
and for crossing odd amplitude ($\xi$=-1)
\begin{equation}\label{eq:CE-part-ampl-xi=-1}
\varphi^-(\omega,t)=-32\sqrt{\pi}(2\omega+3)\dfrac{\Gamma(\omega+3/2)}{\Gamma(\omega+2)}2^\omega \dfrac{\phi^-(\omega,t)}{\sin(\pi \omega/2)}.
\end{equation}
Inverse transformation is 
\begin{equation}
\varphi^\pm(\omega,t)=\int\limits_0^\infty d\zeta e^{-\omega \zeta} F^\pm(z_t,t), \qquad z_t=e^\zeta.
\end{equation}
One can show that in order to have maximal growth of total cross section $\sigma_{tot}(s)\propto \xi^2$ at $s\to \infty$, to have a growing elastic cross section bounded by
$$ \sigma_{el}(s)/\sigma_{tot}(s)\to const \quad \text{at}\quad s\to \infty $$
and to provide the correct analytical properties of amplitude at  $t\approx 0$ necessary to write  the  partial amplitude $\phi(\omega,t)$ in the following form (more details are given in the Appendics, Section \ref{sec:appendicsA})
\begin{equation}\label{eq:omega-t FMO}  
\varphi^\pm(\omega,t )=\binom{i}{-1}\dfrac{\beta^\pm(\omega, t)}{[\omega^2+r_\pm^2q_{\perp}^2]^{3/2}}.
\end{equation}
where $r_\pm$ are some constants, $q_{\perp}^2=-t$ and $\beta(\omega,t)$ has not singularity at $\omega^2+R^2q_{\perp}^2=0$. In fact a choice of the sign in $\phi^-(\omega,t)$ does nor matter because the crossing odd terms contribute to $pp$ and $\bar pp$ amplitude with the opposite signs. In order to have agreement with parametrization and parameters which we used in the papers devoted to analysis of the data at $t=0$, we should replace -1 for for +1 in front of  $\phi^-(\omega,t)$. 

At $\omega =0$, function $\varphi^-(\omega,t)$ has singularity in $t$ if $\beta^-(0,t) \neq 0 $, namely, $\phi^-(0,t)\propto (-t)^{3/2}$. One of arguments against the Maximal Odderon is that  this singularity in partial amplitude means the existing of  massless particle in the model. However as we seen   
above $\varphi^-(\omega,t) $ is not the real physical partial amplitude which is
\begin{equation}\label{eq:phys-odd-partial-amplitude}
\begin{array}{ll}
\phi^-(\omega,t)&=\left [32\sqrt{\pi}(2\omega+3)\dfrac{\Gamma(\omega+3/2)}{\Gamma(\omega+2)}2^\omega \right ]^{-1}\\
&\times \sin(\pi \omega/2)\varphi^-(\omega,t)
\end{array}
\end{equation}
and it  equals to 0 at $\omega=0$ because of $\sin(\pi \omega /2)$ coming from signature factor. 

Now let us suppose that in accordance with the structure of the singularity of $\varphi_\pm(\omega,t)$ at $\omega^2+\omega_{0\pm}^2=0$  ($\omega_{0\pm}^2= R_\pm^2q_{\perp}^2)$ the functions $\beta_\pm(\omega,t)$, depending on $\omega$ through the variable $\kappa_\pm =(\omega^2+\omega_{0\pm}^2) ^{1/2}$,  can be expanded in powers of $\kappa_\pm$
\begin{equation}\label{eq:expansion-phi} 
\varphi^\pm(\omega,t)=\binom{i}{1}\dfrac{\beta_1^{\pm}(t)+\kappa_\pm \beta_2^{\pm}(t)+ \kappa_\pm^2 \beta_3^{\pm}(t)}{\kappa_\pm ^{3}}
\end{equation}

Then making use of the table integrals (see the Section \ref{sec:appendicsA}) we obtain the expressions for $F^\pm (z_t,t)$ which are written in the next Section.

\subsection{Froissaron and Maximal Odderon in ($s, t$)-representation}
\label{sec:5}

At $t=0$ Froissaron and Maximal Odderon have the universal form independently of any extension to $t\neq 0$:
\begin{equation}\label{eq:Ft=0}
F^H(z_t,t=0)=iz[H_1\ln^2(-iz_t)+H_2\ln(-iz_t)+H_3], 
\end{equation}
\begin{equation}\label{eq:MOt=0}
F^{MO}(z_t,t=0)=z[O_1\ln^2(-iz_t)+O_2\ln(-iz_t)+O_3] 
\end{equation}
where $z=2m^2z_t$. At $t=0$ we have  $z_t=(s-2m^2)/(2m^2)$.   

The Froissaron and the Maximal Odderon defined at $t=0$ by above Eqs. (\ref{eq:Ft=0}, \ref{eq:MOt=0}) allow various extensions to analytical $t$-dependences. Probably it is impossible {\it a priory} to choose the best  of them. In the present work we consider an extension of Eqs.~(\ref{eq:Ft=0}, \ref{eq:MOt=0}) in accordance with  Eq. (\ref{eq:expansion-phi}).    

\begin{equation}\label{eq:MN}
\begin{array} {ll}
\dfrac{-1}{iz}F^H(z_t,t)=H_1\zeta^2\dfrac{2J_{1}(r_{+}\tau \zeta)} {r_{+}\tau \zeta }\Phi^{2}_{H,1}(t)  \\
+H_2\zeta\dfrac{\sin(r_{+}\tau \zeta)}{r_+\tau \zeta}\Phi^{2}_{H,2}(t)+H_3J_0(r_{+}\tau \zeta)\Phi^{2}_{H,3}(t) ,\\
\Phi_{H,i}(t)=\exp(b^H_iq_+),\quad i=1.2,3 \\
q_+=2m_{\pi }-\sqrt{4m_{\pi }^2-t} .  
\end{array}
\end{equation}
\begin{equation}\label{eq:MO}
\begin{array}{ll}
\dfrac{1}{z}F^{MO}(z_t,t)=
O_1\zeta^2\dfrac{2J_{1}(r_{-}\tau \zeta)}{r_{-}\tau \zeta}\Phi^{2}_{O,1}(t) \\
+ O_2\zeta \dfrac{\sin(r_{-}\tau \zeta)}{r_{-}\tau \zeta}\Phi^{2}_{O,2}(t)
+O_3J_0(r_{-}\tau \zeta)\Phi^{2}_{O,3}(t),\\ 
\Phi_{O,i}(t)=\exp(b^O_iq_-), \quad i=1,2,3,\\
q_-=3m_{\pi }-\sqrt{9m_{\pi }^2-t}.   
\end{array}
\end{equation}
where $z=2m^2z_t, \quad \zeta=\ln(-iz_t),  \quad \tau=\sqrt{-t/t_0}, \quad   t_0=1\text{GeV}^2$.

Due to the factor  $z$ (instead of $z_t$) the  amplitudes $F^H(z_t,t)$ and $F^{MO}(z_t,t)$ have the required normalization with additional factor $2m^2$.

\section{Comparison of the FMO model with the  data}
\label{sec:6}
We give here the results of the fit to the data in the following region of $s$ and $|t|$. 
\begin{equation}\label{eq:dataregion}
\begin{array}[]{llllll}
\text{for} \quad \sigma_{tot}(s), \rho(s) \quad & \text{at} \quad  5 & \text{GeV} & \leq \sqrt{s} & \leq 13 & \text{TeV}, \\

\text{for} \quad d\sigma(s,t)/dt \quad & \text{at} \quad 9 & \text{GeV} & \leq \sqrt{s} & \leq 13 & \text{TeV}\\ 
\nonumber
\text{and}  & \text{at}  \quad 10^{-4}  & \text{GeV}^2 & \leq |t| & \leq  5 & \text{GeV}^2.
\end{array}
\end{equation}
We add the recent data at $t=0$ of  TOTEM Collaboration \cite{TOTEM-01, TOTEM-02, TOTEM-03, TOTEM-04} to  data set published by Particle Data Group \cite{PDG}.

We have performed two alternative fits of the FMO model and experimental data from the above mentioned kinematic region.

{\bf In the Fit I} we take into account all the data at t=0, i.e. $\sigma_{tot}$ and $\rho$  are calculated from the FMO model,  free parameters are determined from the fit to all $t$, chosen in such a way that we can  ignore in given region  the contribution of the Coulomb part of amplitudes which is less than ~1\% of the nuclear amplitude. Thus, $t$-region $0.0<|t|<0.05$ GeV$^2$ is excluded in the Fit I, and the Coulomb part of amplitudes put to zero.  

{\bf In the Fit II } all experimental data on $\sigma_{tot}$ and $\rho$ are excluded and fit is made at energies $\sqrt{s}> 19$ GeV  and  $0<|t|< 5 $ GeV$^2$. Taking into  account that in this kinematic region parameters of CE and CO secondary reggeons are badly determined, we put all the parameters of these contributions as fixed from the results of Fit I.   

For 13 TeV TOTEM data we used the data at $t=0$ for  $\sigma_{tot}$ \cite {TOTEM-02} and $\rho$ \cite{TOTEM-01}, as well as the data on differential cross sections \cite{TOTEM-1, TOTEM-2, Nemes-2018, Ravera-2018}. We add also recently published data on  $d\sigma/dt$ at $\sqrt{s}=2.76$  TeV obtained by TOTEM \cite{TOTEM-3}. 

\subsection{Coulomb amplitude, one of the simplest parameterizations}

Coulomb terms in the $pp$ and $\bar pp$ amplitudes are written in the well known form 
\begin{equation}
{\cal F}_{CN}(s,t)=\pm 8\pi s\dfrac{\alpha}{t}F_1^2(t)\exp(i\alpha \phi(s,t)) 
\end{equation}
where  $\alpha =7.297352 \times 10^{-3} = 1/137.035$ is the fine-structure constant and
\begin{equation}
\begin{array}{rl}
F_1(t)&=\dfrac{4m_p^2-\mu_pt}{4m_p^2-t}\dfrac{1}{(1-t/0.71)^2},\\
&\\
\mu_p& =2.7928473446 
\end{array}
\end{equation}
where $\mu_p$ is the magnetic momemtum of proton. For the phase $\phi(s,t)$ we nave
\begin{equation}
\phi(s,t)=\pm \left [ \ln \left ( \dfrac{B(s)}{2}|t|\right )+\gamma \right]
\end{equation}
where $\gamma=0.5772156649$ is the Euler constant.

The slope $B(s)$ is calculated through a fit making use of the equation
\begin{equation}\label{eq:averagedB}
\begin{array}{ll}
<B(s)>&=(1/\Delta_t)\int \limits _{t_{max}}^{t_{min}}dt \,\dfrac{d}{dt}(\ln(d\sigma(t)/dt))\\
&= (1/\Delta_t )\left  [\ln\left (\dfrac{d\sigma(t_{min})/dt}{d\sigma(t_{max})/dt}\right )\right ]
\end{array}
\end{equation}
where $\Delta_t=t_{max}-t_{min}$. We put (in accordance with the TOTEM estimations \cite{TOTEM-2}), $t_{max}=-0.07$ GeV$^2$,  $t_{min}=-0.005$ GeV$^2$.

%
\begin{figure*} 
	\includegraphics{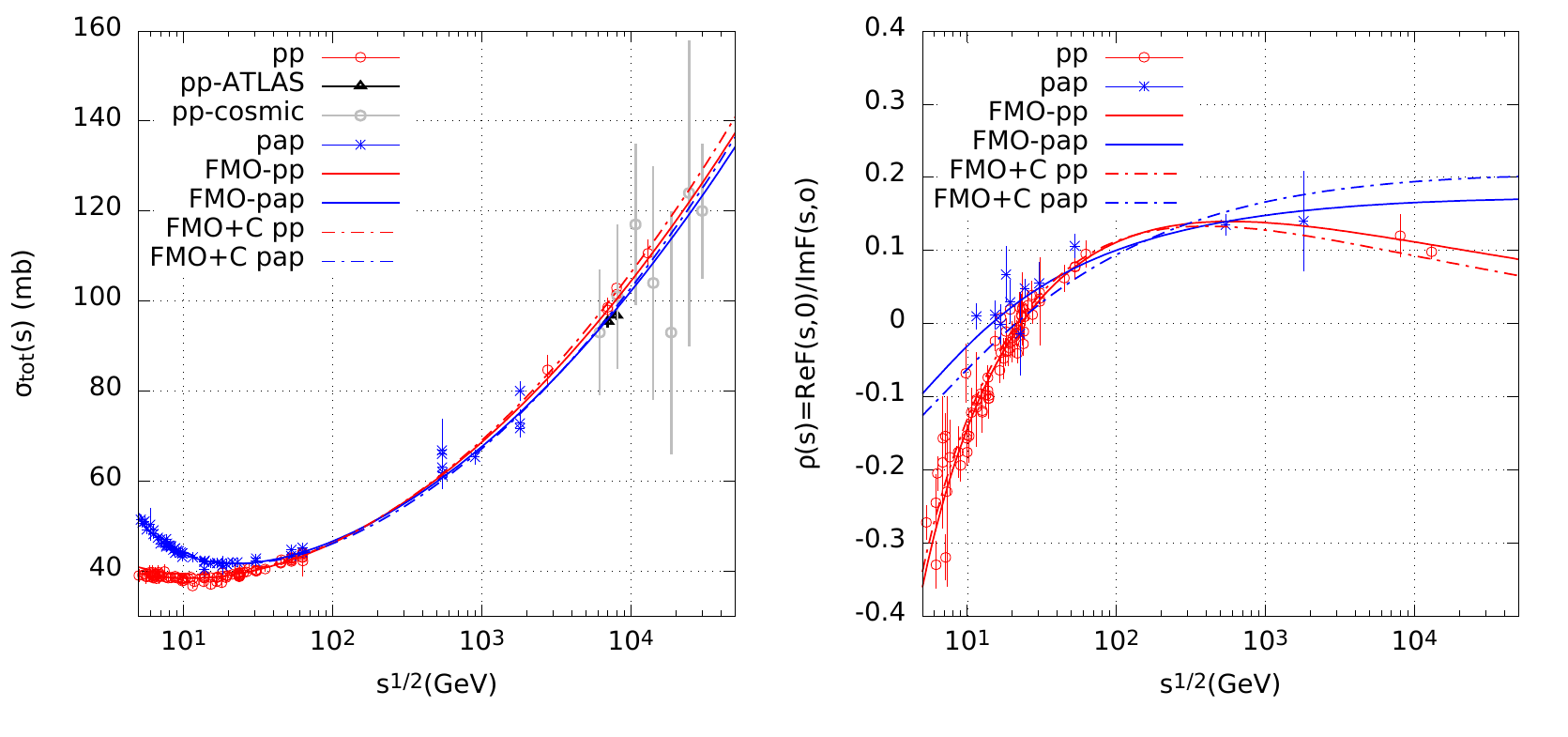}
	\caption{Total cross sections and ratios $\rho$ in FMO model with the $PP,PO, OO$ terms added}
	\label{fig:sig-tot-and-rho}       
\end{figure*}
%
%

\subsection{$pp$ and $\bar pp$ differential cross sections  $d\sigma/dt$}
\label{sec:9}
\begin{figure}[H]
	\centering
	\resizebox{0.5\textwidth}{!}{%
		\includegraphics{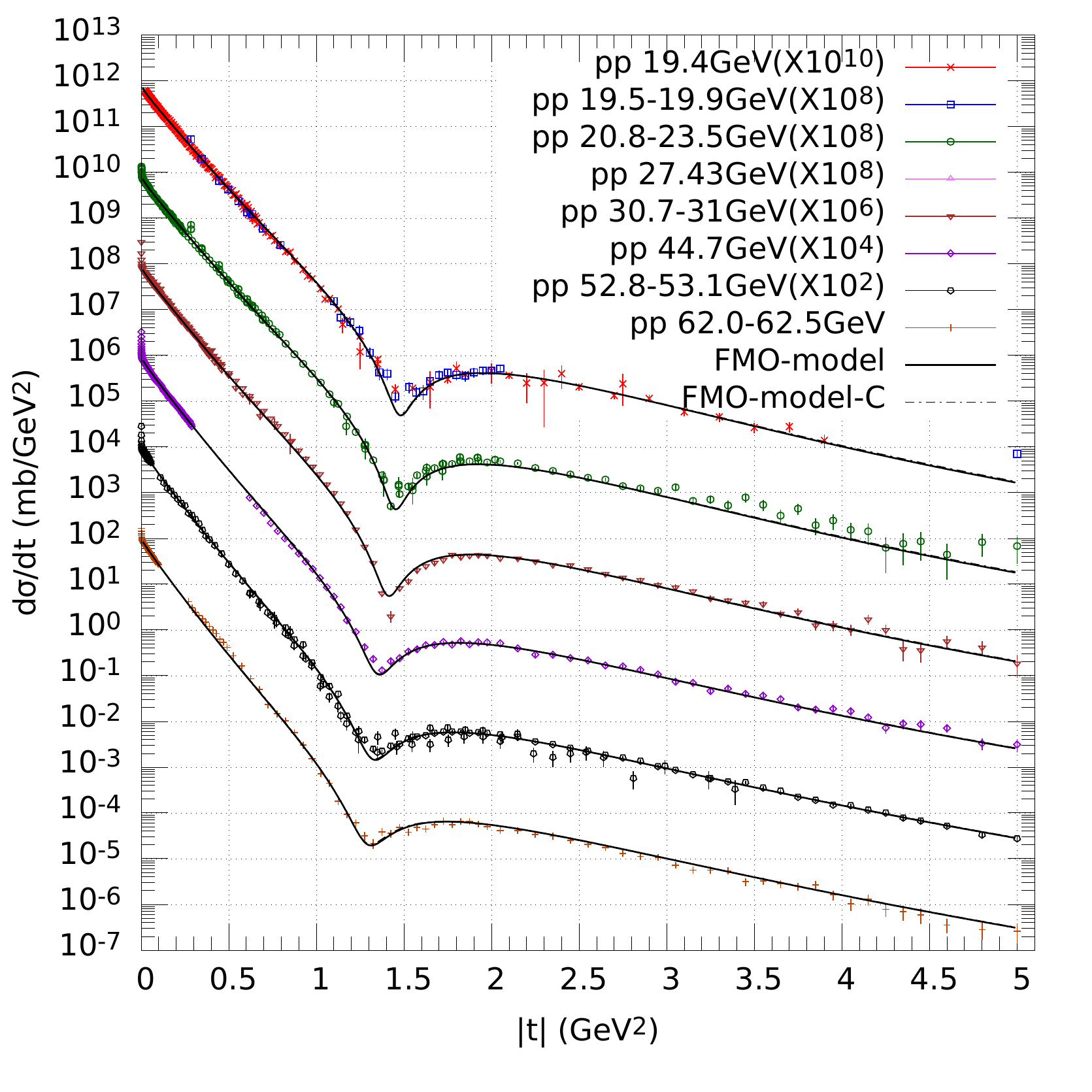}
	}
	\caption{$pp$ differential cross sections at $\sqrt{s}>19  $ GeV}
	\label{fig:pp-high-fmo}
\end{figure}

\begin{figure}
	\centering
	\resizebox{0.5\textwidth}{!}{%
\includegraphics{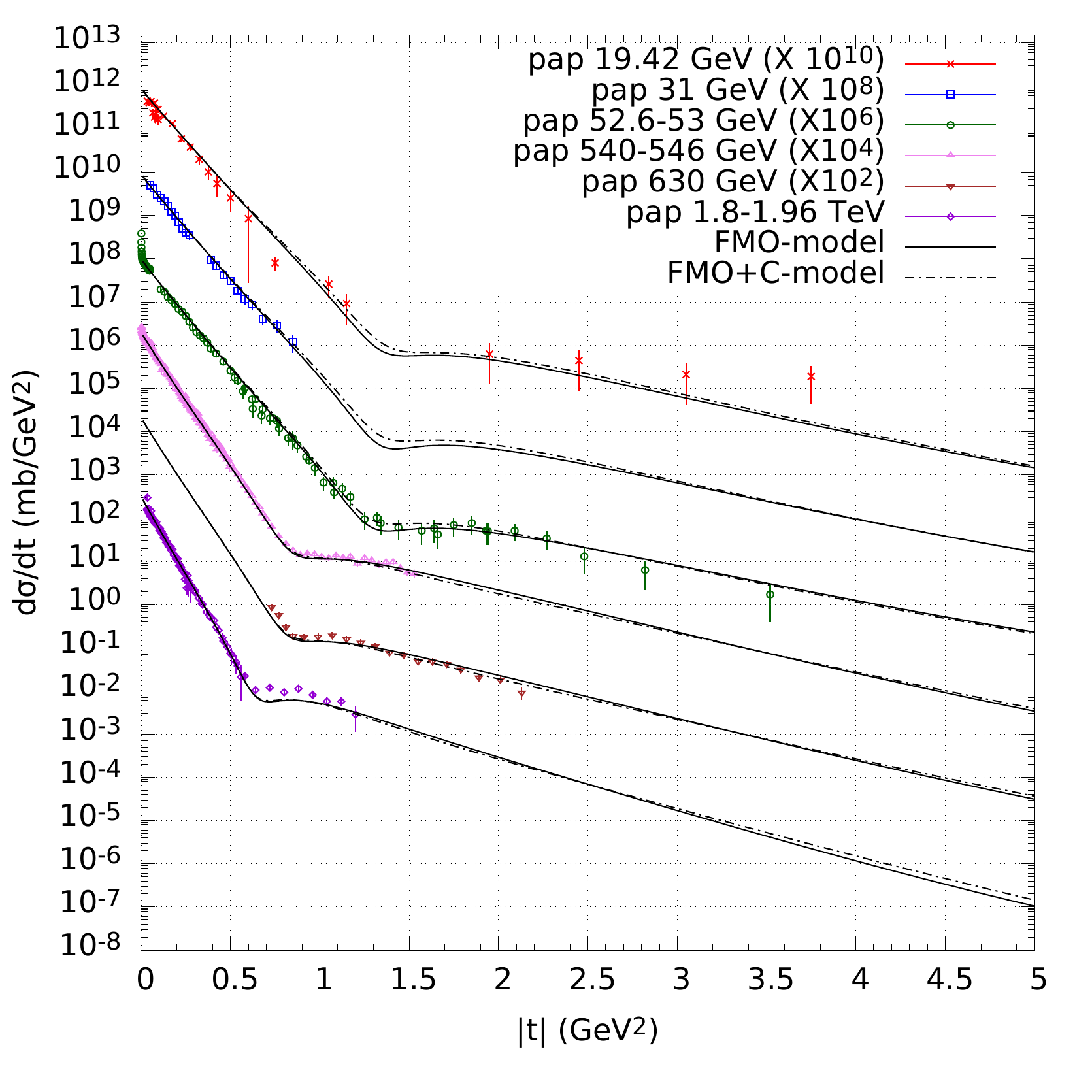}
	}
 	\caption{$\bar pp$ differential cross sections at $\sqrt{s}$  from 19 GeV  up to 1.96 TeV}
	\label{fig:pap-dsdt-high-fmo}
\end{figure}

\begin{figure}[!hbp]
	\centering
	\resizebox{0.5\textwidth}{!}{%
	\includegraphics{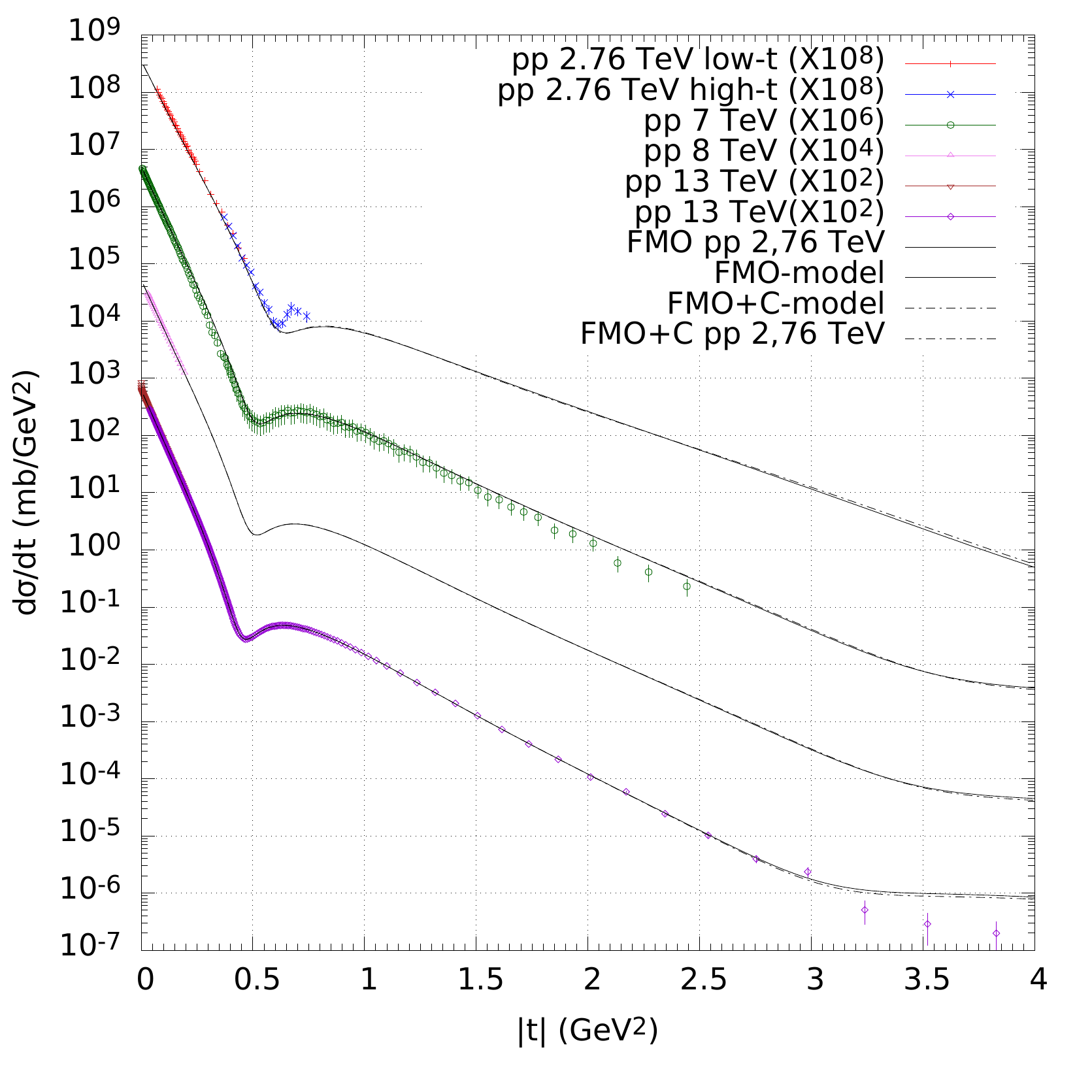}
	}
	\caption{$pp$ differential cross sections at $\sqrt{s}=7, 8, 13 $ TeV}
	\label{fig:pp-dsdt-totem-fmo}
\end{figure}

\begin{figure}[!hbp]
	\centering
	\resizebox{0.5\textwidth}{!}{%
	\includegraphics{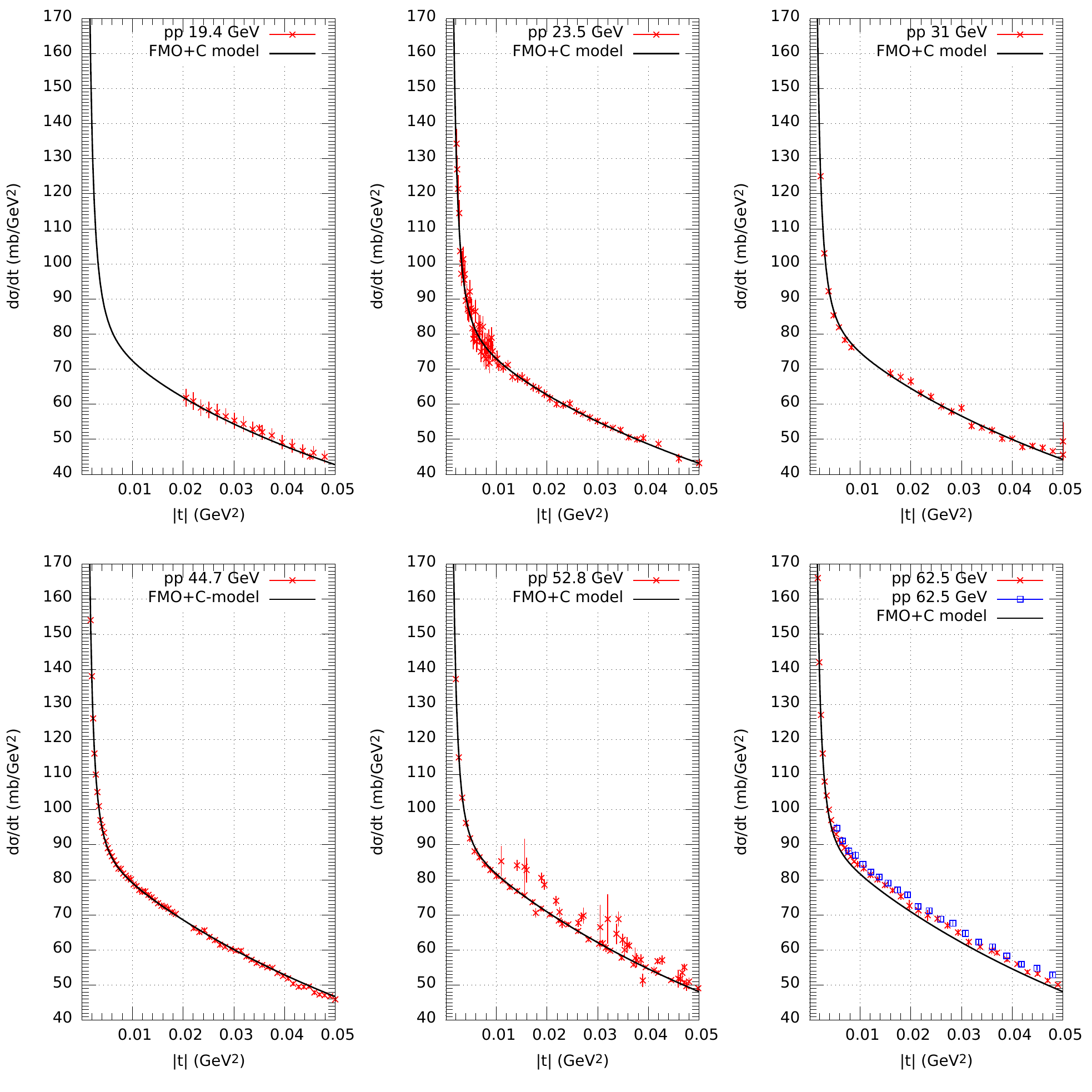}
	}
	\caption{Differential $pp$ cross sections at the lowest $|t|$ and at ISR energies}
	\label{fig:pp-fmo-ISR-low-t-C}
\end{figure}

\begin{figure}[!hbp]
	\centering
	\resizebox{0.5\textwidth}{!}{%
		\includegraphics{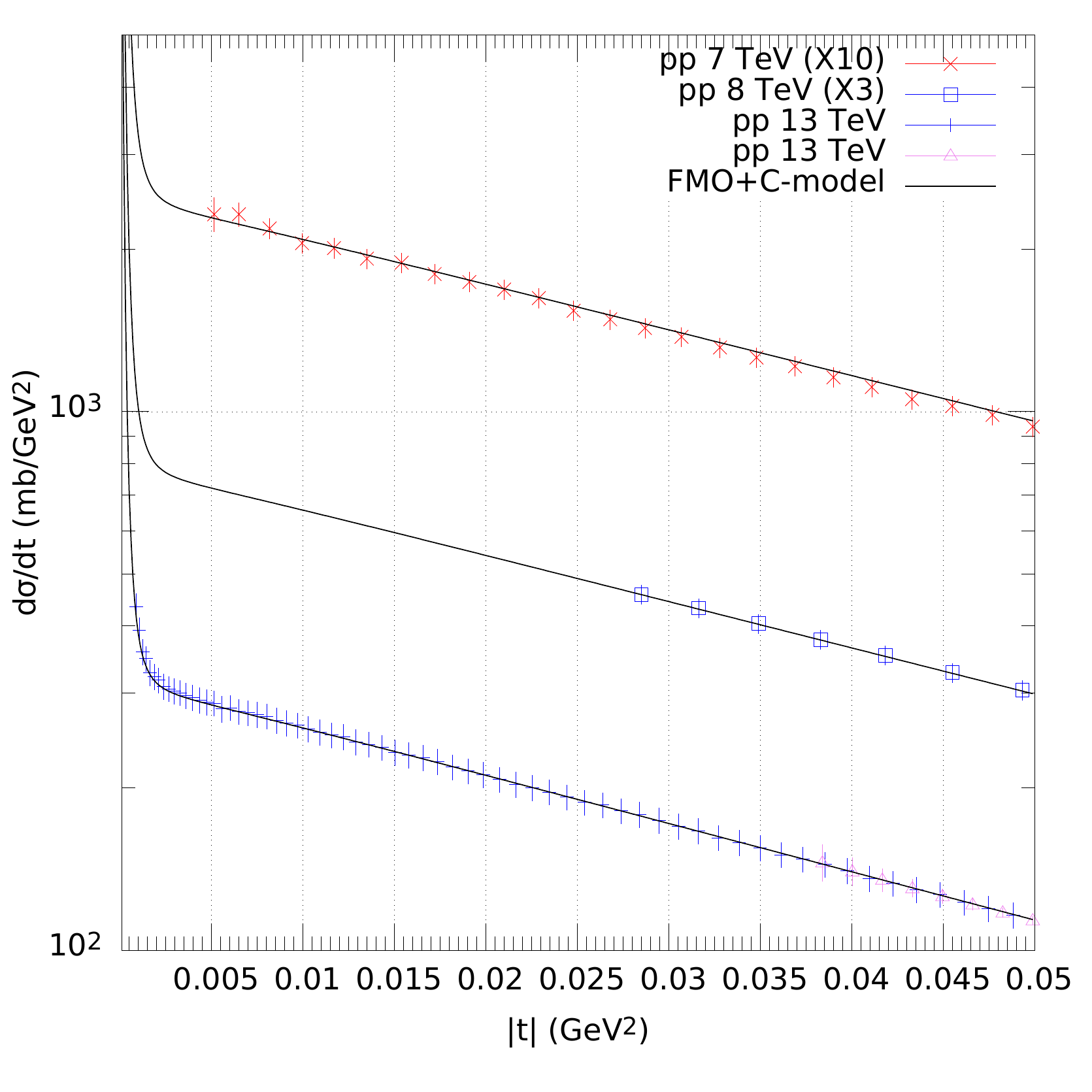}
	}
	\caption{Differential $pp$ cross sections at the lowest $|t|$ and at LHC energies}
	\label{fig:pp-fmo-LHC-low-t-C}
\end{figure}

\begin{figure}[!hbp]
	\centering
	\resizebox{0.5\textwidth}{!}{%
		\includegraphics{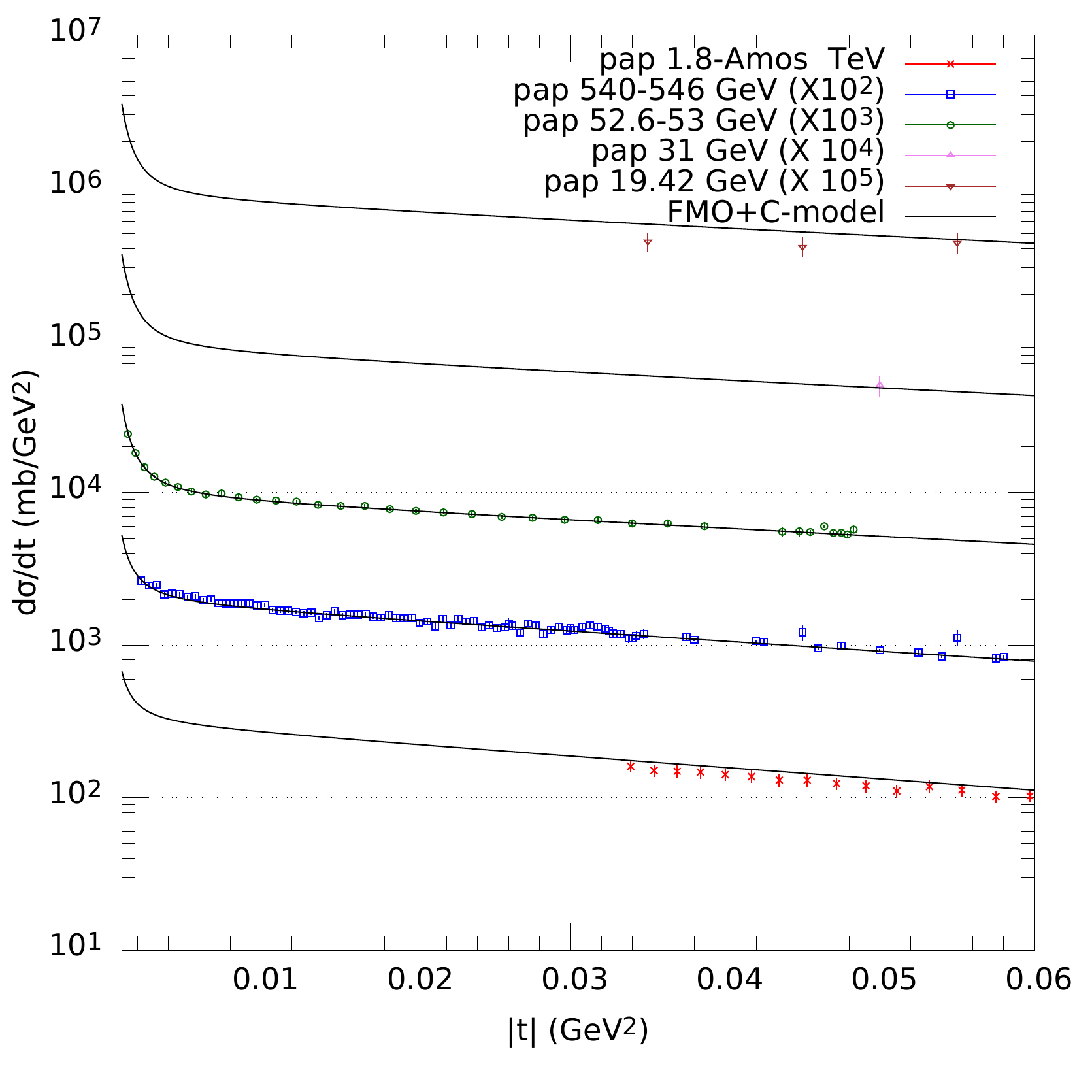}
	}
	\caption{Differential $\bar pp$ cross sections at the lowest $|t|$}
	\label{fig:pap-fmo-low-t-C}
\end{figure}

\begin{figure}[!hbp]
	\centering
	\resizebox{0.5\textwidth}{!}{%
	\includegraphics{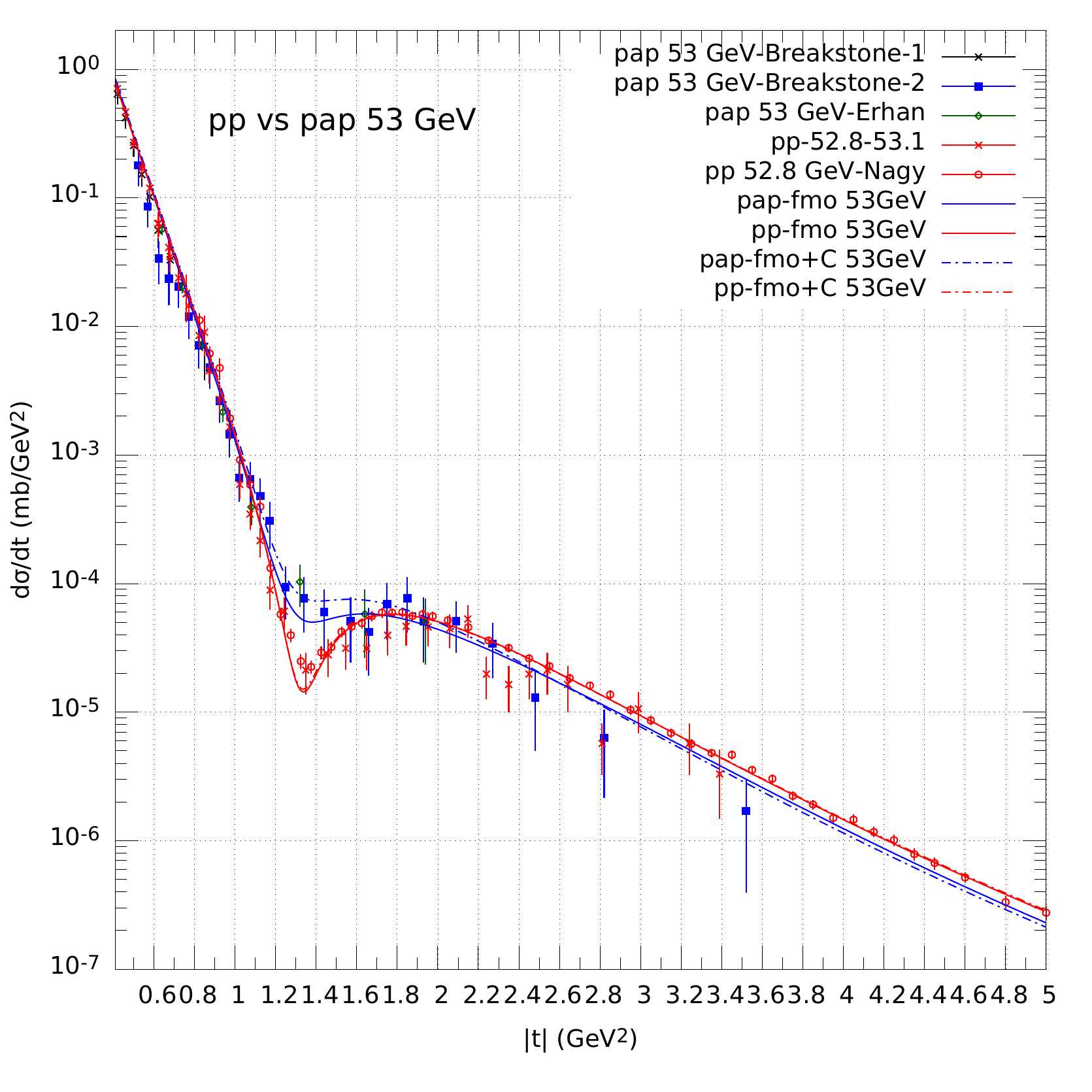}
	}
	\caption{$pp$ and $\bar pp$ differential cross sections at $\sqrt{s}=53$ GeV}
	\label{fig:pp-pap-53gev-fmo}
\end{figure}

\begin{figure}[!hbp]
	\centering
	\resizebox{0.5\textwidth}{!}{%
	\includegraphics{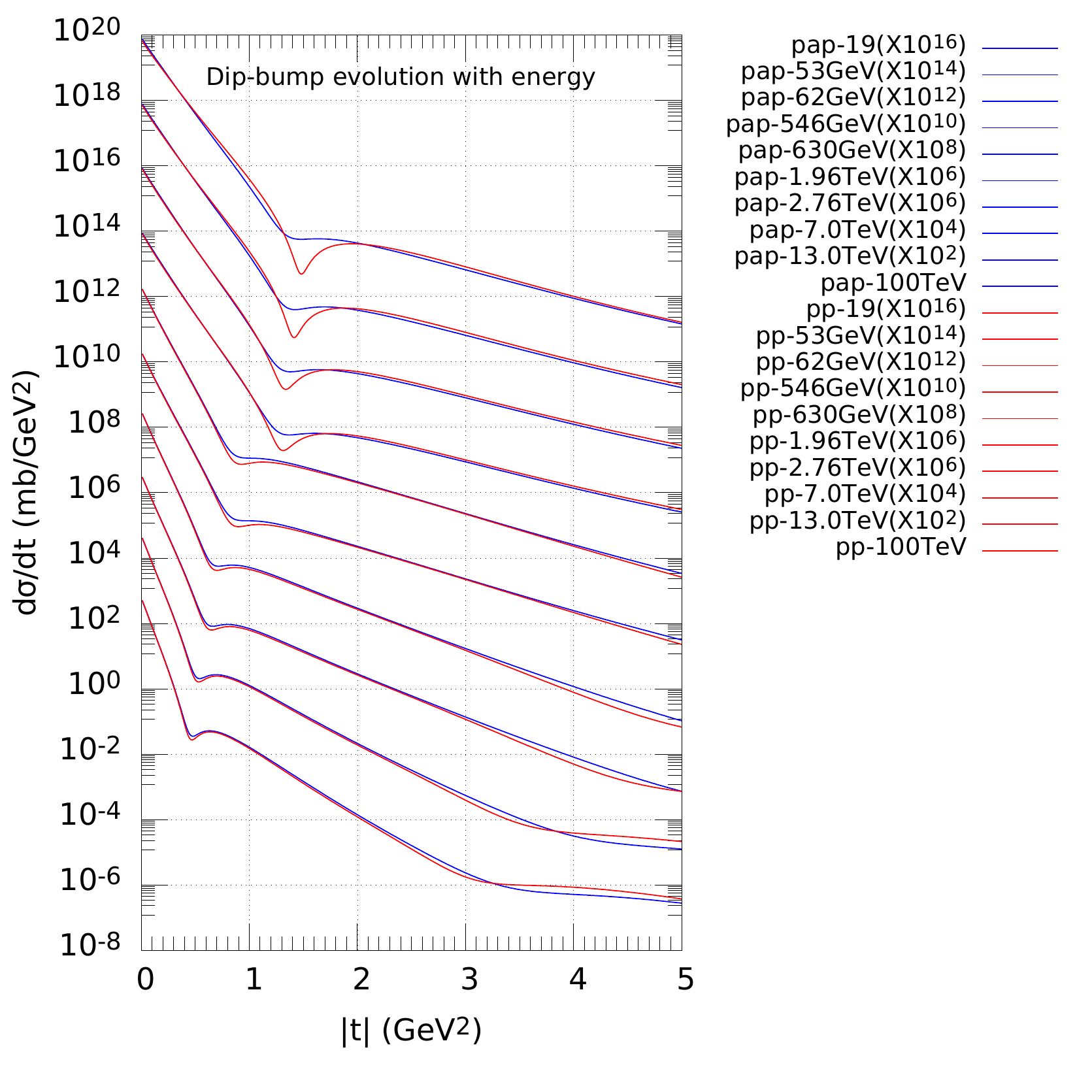}	
}
	\caption{Evolution of $pp$ and $\bar pp$ differential cross sections with increasing energy}
	\label{fig:pap-dsdt-all-fmo}
\end{figure}

\begin{figure}[!hbp]
	\centering
	\resizebox{0.5\textwidth}{!}{%
	\includegraphics{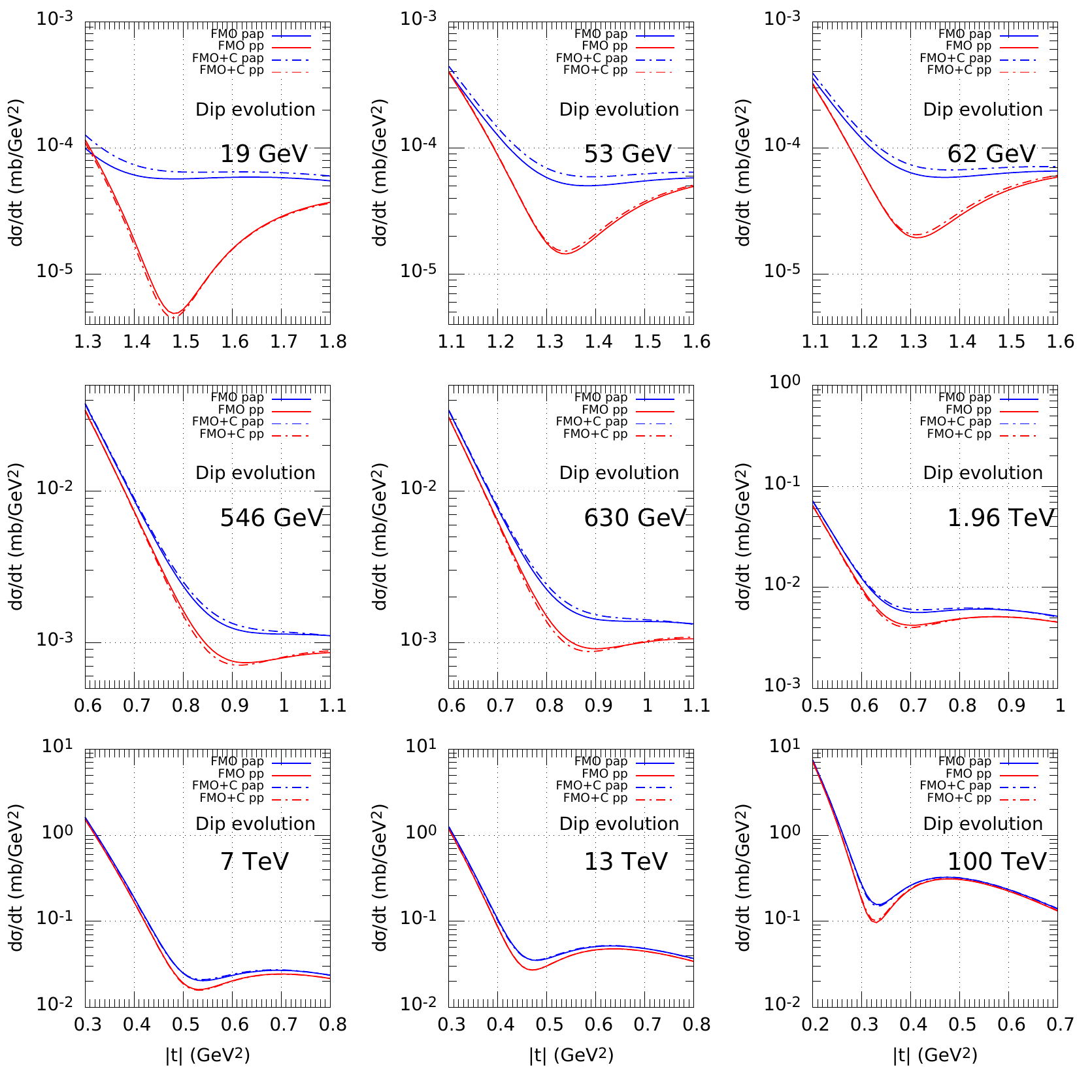}	
}
	\caption{$pp$ and $\bar pp$ differential cross sections in and around the dip region}
	\label{fig:dsdt-multi-fmo}
\end{figure}

\begin{figure}[!hbp]
	\centering
	\resizebox{0.5\textwidth}{!}{%
	\includegraphics{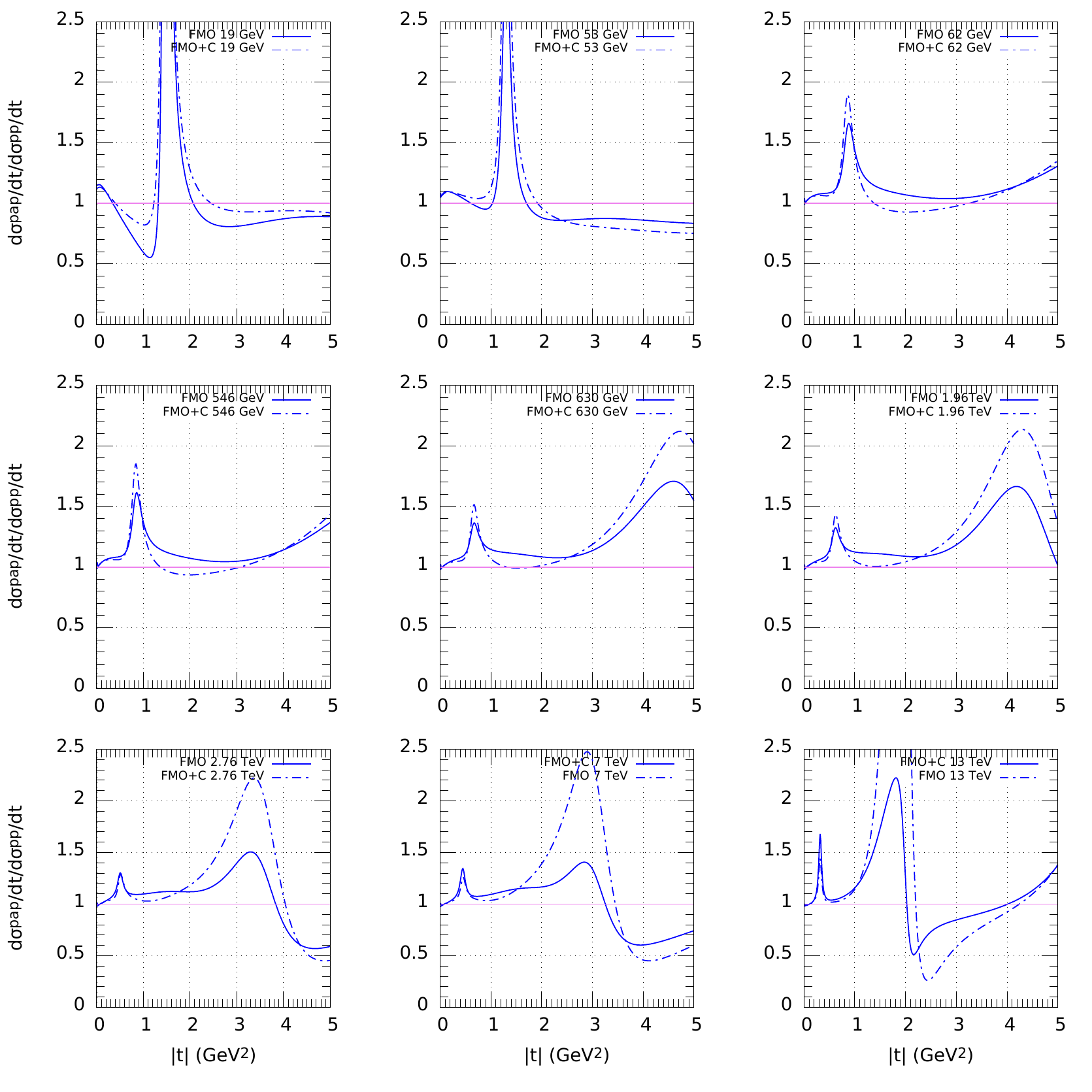}
	}
	\caption{Evolution of the ratio of differential cross sections $R_{\sigma }=(d\sigma(\bar pp)/dt)/(d\sigma(pp)/dt)$ with energy} 
	\label{fig:delta-dsdt-fmo}
\end{figure}

\begin{figure}
	\centering
	\includegraphics[width=1.0\linewidth]{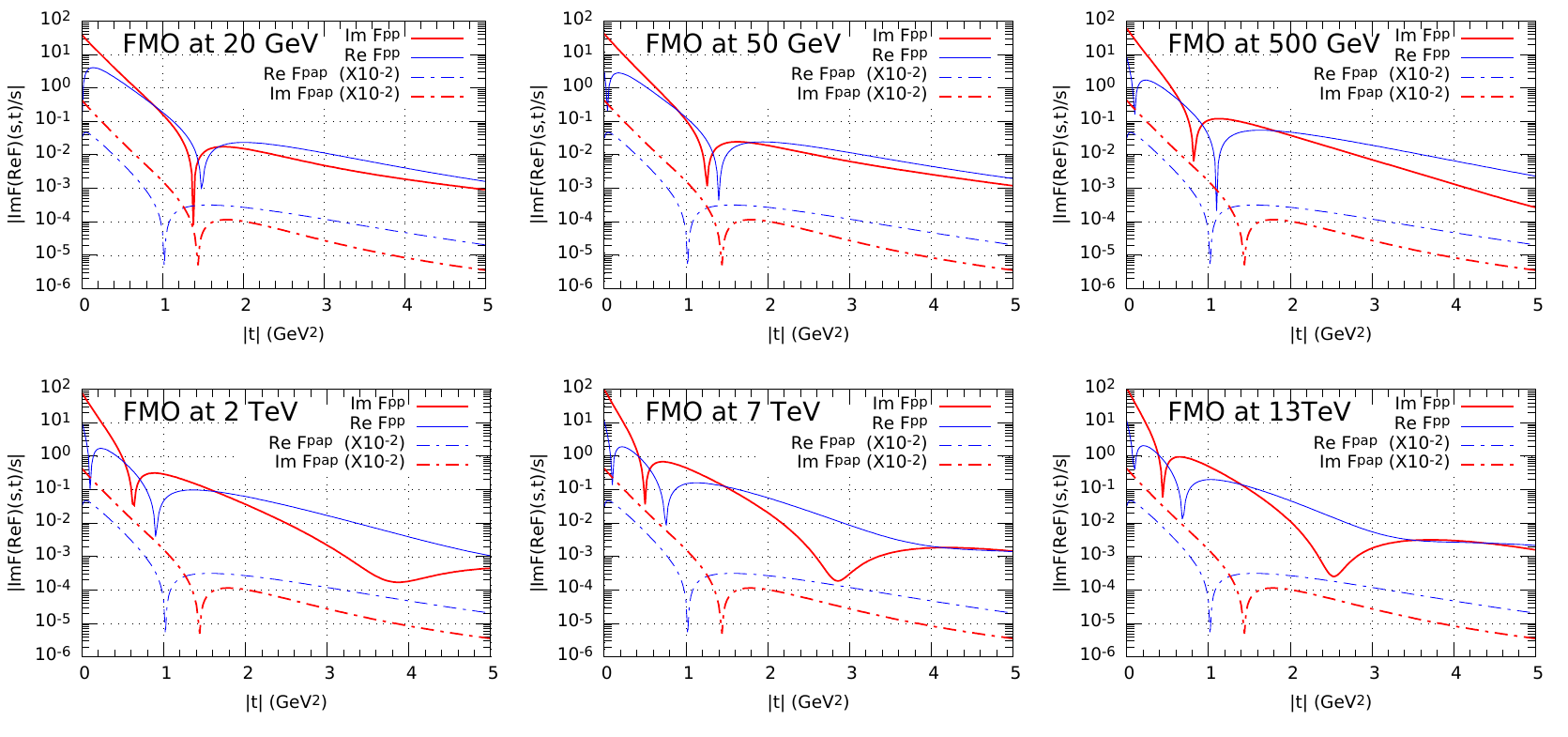}
	\caption{Partial contributions of the real and imaginary parts of even and odd terms to $pp$ and $\bar pp$ scattering amplitudes at various energies}
	\label{fig:fmo-partsfull-20g-13t}
\end{figure}

\begin{figure*}
	\includegraphics{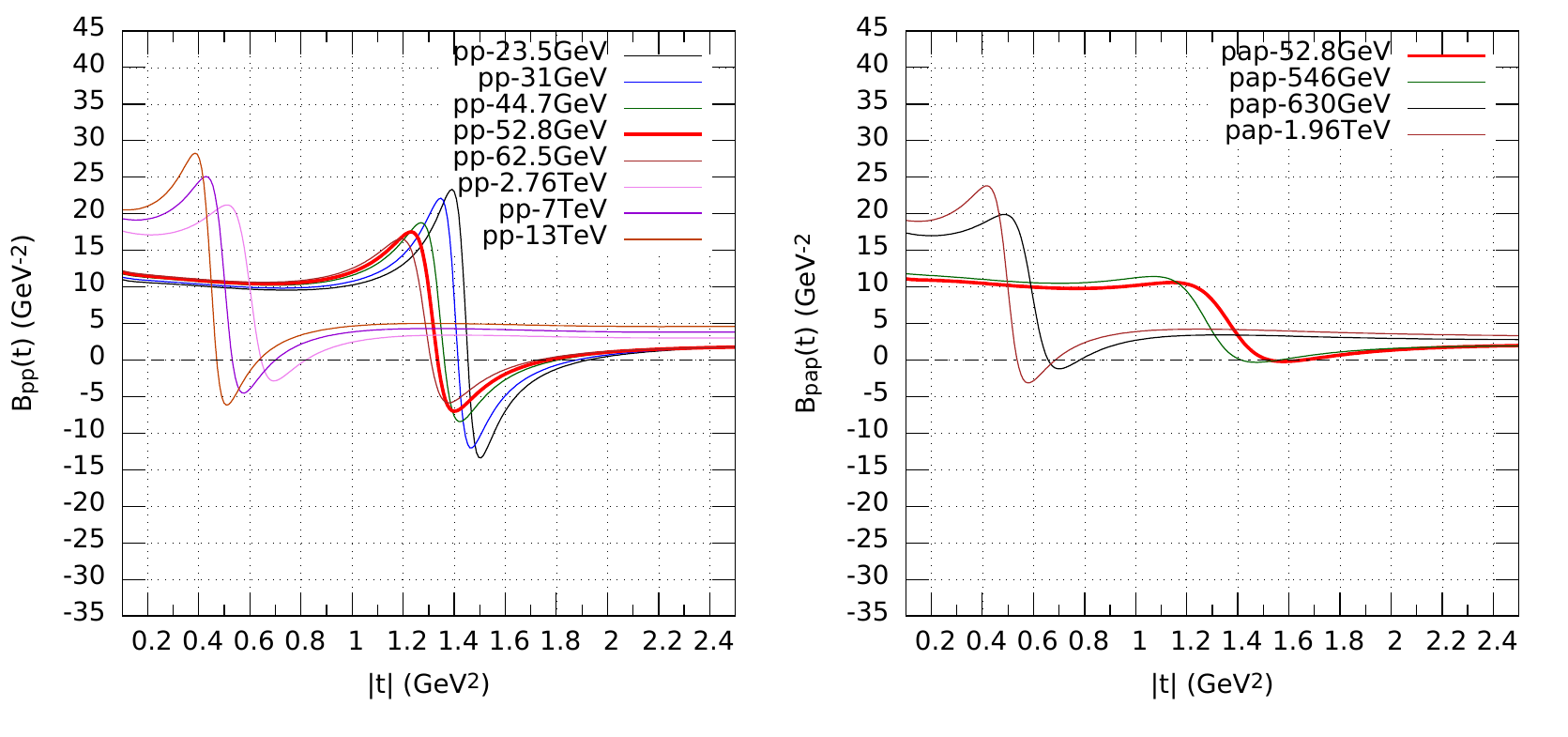}
	\caption{Slope $B(s,t)$ for $pp$ (left panel) and $\bar pp)$ (right panel) at selected energies}
	\label{fig:B(s_exp,t)}       
\end{figure*}

\begin{figure}
	\centering
	\resizebox{0.5\textwidth}{!}{%
		\includegraphics{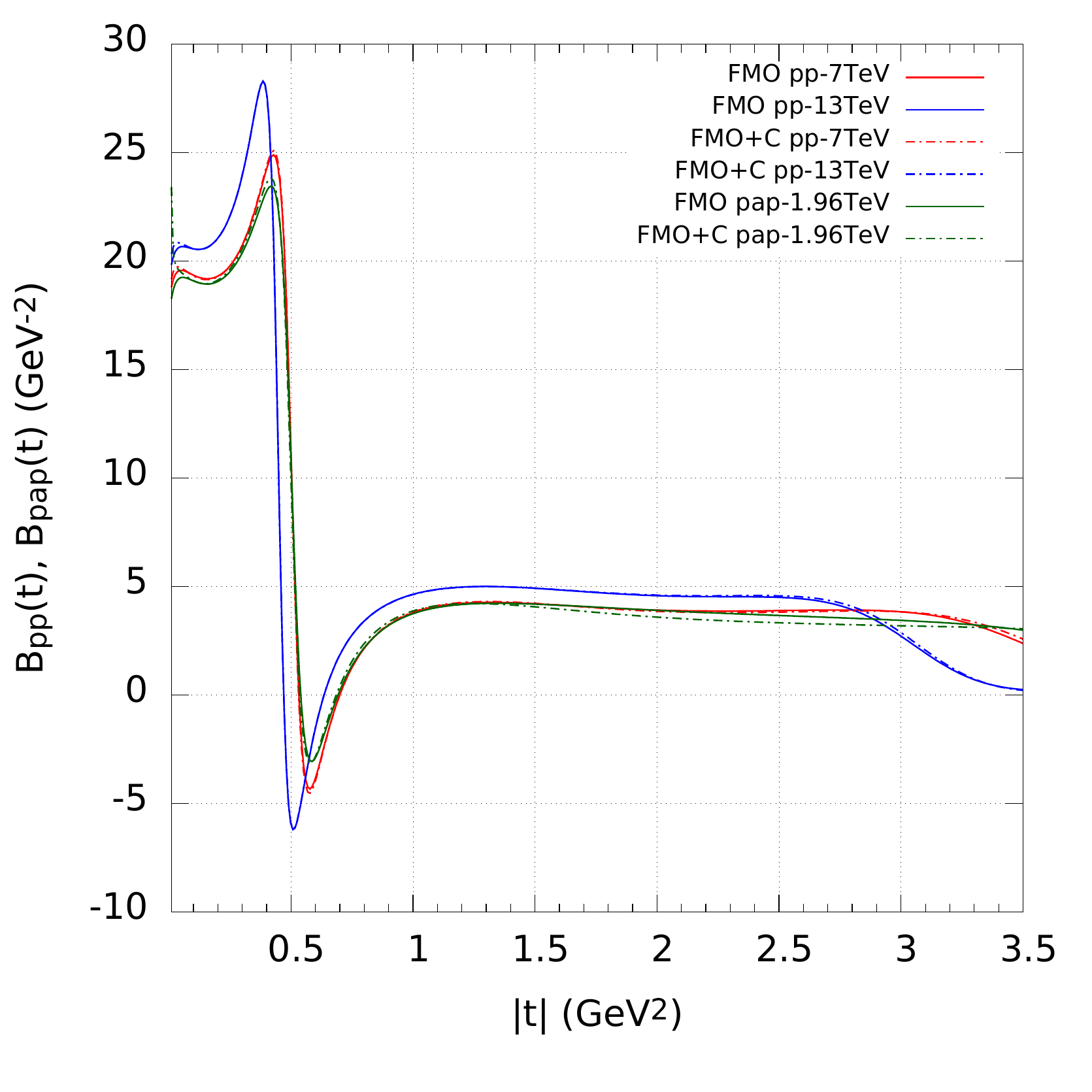}
	}
	\caption{Dependence on $t$ of the slopes $B(s,t)$  for $pp$ scattering at 7 and 13 TeV and
for  $\bar pp$  scattering at 1.96 TeV
}
	\label{fig:B(s,t)7-13pp,1_96pap}
\end{figure}

\begin{figure}[!hbp]
	\centering
	\resizebox{0.5\textwidth}{!}{%
		\includegraphics{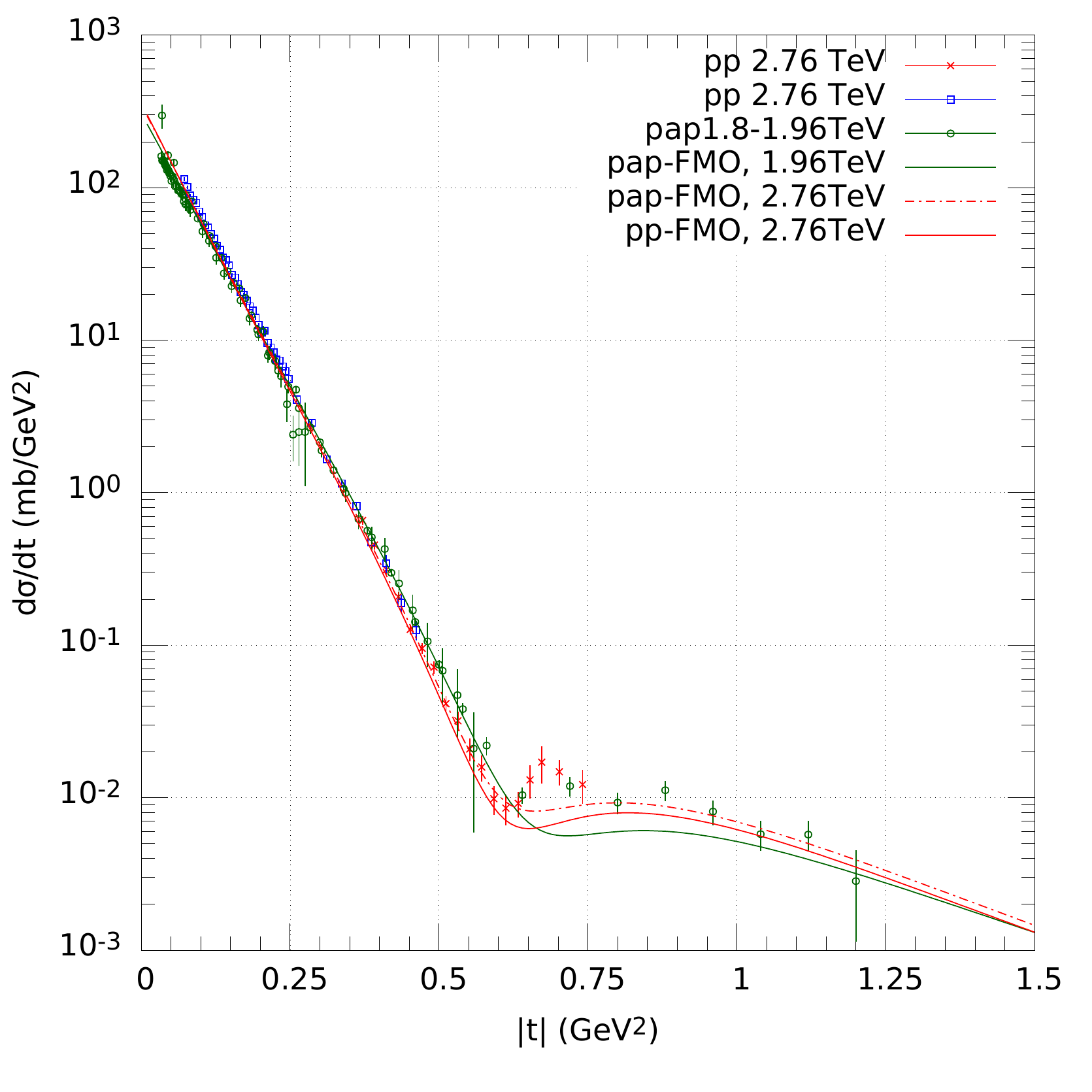}
	}
	\caption{$\bar pp$ differential cross section at 1.18-1.96 TeV and $pp$ differential cross section at 2.76 TeV}
	\label{fig:pp-2.76-pap-1.8-FMO}
\end{figure}

Here we present results for both methods of the data description. {\bf Fit I}: FMO model without Coulumb term fitted to the  whole set of data excluding lowest $|t|<0.05$ GeV$^2$.  {\bf Fit II}: FMO model with Coulomb term fitted to the whole set of the data at $t\neq 0$. In the legends of Fig. \ref{fig:sig-tot-and-rho}-\ref{fig:fmo-partsfull-20g-13t}  these fits are labeled as "FMO"  and "FMO+C", correspondingly. The  curves shown at the Figs. \ref{fig:B(s)B(t)multi} - \ref{fig:pp-2.76-pap-1.8-FMO} were calculated in the FMO model without Coulomb terms ({\bf Fit I}).

Number of experimental points in $pp$ and $\bar pp$ total cross sections $\sigma_t^{pp},  \sigma_t^{\bar pp}$, ratios  $\rho^{pp}, \rho^{\bar pp}$ and differential cross sections used in the {\bf Fit I}  and quality of fit are shown in the Table 1. 

Numbers of the data points and obtained values of $\chi{^2}$ in the {\bf Fit II}  are given in the Table 2.

\begin{table}[!hbp]
	\label{tab:observables-0}
	\centering
	\begin{tabular}{cccc}
				\multicolumn{4}{c}{FMO without Coulomb terms}\\
		\hline
		Process   &     Observable    &    N, number     & $\chi^2/N$   \\ 
		&              &    of data     &    \\ 
		\hline 
		$pp$          &     $\sigma_{tot}$    &     110     &   0.857213E+00  \\
		$\bar pp$   &     $\sigma_{tot}$    &      59     &   0.992282E+00  \\
		$pp$          &       $\rho$               &       67      &  0.169032E+01  \\
		$\bar pp$   &       $\rho$              &       12       &  0.836012E+00  \\
		$pp$          &      $d\sigma/dt$     &  1574        &    0.174594E+01    \\
		$\bar pp$   &      $d\sigma/dt$     &   389       &     0.121600E+01    \\
		\hline
		\multicolumn{4}{l}{ $\chi^2_{tot} =3718.994$}         $\chi^2/\text{NDF}=1.613$   \\ 
	\end{tabular} 
	\caption{Number of experimental points and the quality of their description when the usual  minimization in FMO model is applied}
\end{table}

\begin{table}[!hbp]
	\label{tab:observables-1}
	\centering
	\begin{tabular}{cccc}
				\multicolumn{4}{c}{FMO with Coulomb terms}\\
				\hline
		Process   &     Observable    &    N, number     & $\chi^2/N$   \\ 
		&              &    of data     &    \\ 
		\hline 
		$pp$          &      $d\sigma/dt$     &    2492        & 0.164888E+01   \\
		$\bar pp$   &      $d\sigma/dt$     &    536           &  0.121288E+01   \\
		\hline
		\multicolumn{4}{l}{$\chi^2_{tot}$= 4790.652 }        $\chi^2/\text{NDF}$=1.584 
	\end{tabular} 
	\caption{Number of experimental points and the quality of their description when the fit with FMO+Coulomb terms is made. The data  on $\sigma_{tot}(s)$ and $\rho(s)$ has been excluded from this fit}
\end{table}

The values of parameters and their errors obtained in these two fits  within the  FMO model are given in the Table 3 (parameters of the Froissaron and Maximal Odderon terms,  of the standard Pomeron and Odderon, of the ''hard``Pomeron and Odderon, and of the secondary reggeons). 

To avoid a possible negative cross sections in the large partial waves, $j$, (at the edge of the disk) we put in the fit the restriction $r_-\leq r_+$. However, we observed that in the various considered modifications of the FMO model these parameters are almost equal each to other. Based on this fact we put  $r_-= r_+$ in the model presented here. Also we have fixed  the parameters $b^{\pm}$ at 0 because in all considered fits $b^+$ has the error comparable with the value of parameter and $b^-$ has value close to 0. 

Fig. \ref{fig:sig-tot-and-rho}  demonstrates  a behavior of the $pp$ and $\bar pp$ total cross  sections and ratios real part to imaginary part  of the amplitudes at $t=0$ obtained in the both {\bf Fit I} and {\bf Fit II}. We would like to notice the interesting odderon effect: the change of sign in the differences between total cross section and $\rho$’s between $\sqrt{s}\approx 50$ and $\sqrt{s}\approx 500$ GeV. Such a spectacular effect is allowed by asymptotic theorems. A detailed dynamic model for this effect was not yet invented.

In Figs. \ref{fig:pp-high-fmo} and \ref{fig:pap-dsdt-high-fmo} we show the differential cross-sections at energies bigger than 19 GeV. In  Fig. \ref{fig:pp-dsdt-totem-fmo} we show the differential cross-sections at  the  LHC energies 7, 8 and 13 TeV and in Figs. \ref{fig:pp-fmo-ISR-low-t-C}, \ref{fig:pp-fmo-LHC-low-t-C}, \ref{fig:pap-fmo-low-t-C} we show differential $pp$ and $\bar pp$ at lowest $|t|$.  In Fig. \ref{fig:pp-pap-53gev-fmo},  we show in a magnified way  the differential cross-sections at 53 GeV.

As  one can see from these figures our description of the data in a wide range of ener\-gies  is very good. In Fig.~\ref{fig:pap-dsdt-all-fmo} we show the evolution of  the dip-bump structure  in $pp$ and $\bar pp$ differential cross  sections with increasing energy.  In Fig. \ref{fig:dsdt-multi-fmo} we show in a magnified  way the dip-bump  region at different energies and in Fig.~\ref{fig:delta-dsdt-fmo} we show the evolution of the ratio $R_\sigma = (d\sigma(\bar pp) /dt) / (d\sigma(pp)/dt)$ with increasing energy.  A remarkable prediction can be seen from these last three figures: the difference in the dip-bump region between $\bar pp$  and  $pp$ differential cross sections  is diminishing with increasing energies and, for very high energies (say 100 TeV, see Fig. \ref{fig:dsdt-multi-fmo}), the ratio in the dip-bump region goes to 1. At ISR energies until $\sim 60$  GeV the ratio $R_\sigma > 1$ and then it  becomes less than 1 but increases to maximum at some $t_m$. After  maximum the value of  $R_\sigma$ is decreasing and equals to 1 at some $t_1$ which is going to lower $t$ with increasing energy. At higher $t$ however $R_\sigma$ is oscillating around of 1 when $t$ increases. This is a spectacular Odderon effect.  One can see  also the clear Odderon effects and their evolution with energy in Fig. \ref{fig:fmo-partsfull-20g-13t}.

\begin{table*}[!bhp]
	\label{tab:6}
	\begin{tabular}{lcccc}
		\hline
		                                      &                    \multicolumn{4}{c}{FMO model}                    \\ \hline
		                                      &   \multicolumn{2}{c}{Minimization  without}   &   \multicolumn{2}{c}{Minimization with}   \\
		                                      & \multicolumn{2}{c}{Coulomb terms} & \multicolumn{2}{c}{Coulomb terms } \\ \hline
		Name (dimension)                      &  Value  	&      Error         & Value  	&          Error         \\
\hline
$\alpha_{P}'(\text{GeV}:2)$ &  0.18845E+00 & 0.15606E-03 & 0.16274E+00 &  0.12999E-03\\
$C^{P}$  (mb)                 &  0.67305E+02 & 0.50925E-01 & 0.67098E+02 &  0.40739E-01\\
$b^{P}_{1} (\text{GeV}^2)$       &  0.57234E+01 & 0.54856E-02 & 0.60451E+01 &  0.48533E-02\\
$d_{p}$                            &  0.69294E+00 & 0.71856E-03 & 0.68457E+00 &  0.57001E-03\\
$b^{P}_{2} (\text{GeV}^2)$       &  0.23392E+01 & 0.31028E-02 & 0.24359E+01 &  0.26420E-02\\
\hline                                                                                        
$C^{HP}$   (mb)                     & -0.61825E+02 & 0.35377E-01 & 0.69984E+02 &  0.30999E-01\\
$t_{HP} (\text{GeV}^2)$            &  0.41803E+00 & 0.14780E-03 & 0.41377E+00 &  0.12323E-03\\
\hline                                                                                        
$\alpha_{O}^{'}(\text{GeV}^{-2})$ &  0.15673E-01 & 0.11401E-03  & 0.12298E-01  &  0.10018E-03\\
$C^{O}$ (mb)                          &  0.29156E+02 & 0.25668E-01 & 0.31654E+02 &  0.24212E-01\\
$b^{O}_{1}  (\text{GeV}^{-2})$              &  0.50899E+01 & 0.46679E-02 & 0.52749E+01 &  0.39244E-02\\
$d_{o}$                                   &  0.74110E+00 & 0.65773E-03 & 0.76966E+00 &  0.51463E-03\\
$b^{O}_{2} (\text{GeV}^{-2})$       &  0.21098E+01 & 0.22047E-02 & 0.20931E+01 &  0.21404E-02\\
\hline                                                                                         
$C^{HO}$   (mb)                     &  0.37930E+02 & 0.37256E-01 & 0.42175E+02 &  0.36723E-01\\
$t_{HO} (\text{GeV}^2)$                    &  0.58624E+00 & 0.36266E-03 & 0.55774E+00 &  0.30146E-03\\
\hline                                                                                         
$\alpha_{+}(0)$                       &  0.47754E+00 & 0.51446E-02 & 0.47754E+00 & fixed              \\
$\alpha_{+}^{'} (\text{GeV}^{-2})$ &  0.80000E+00 & 0.31788E-02 & 0.80000E+00 &  fixed              \\
$C^{+}$  (mb)                          &  0.47341E+02 & 0.11590E+01 & 0.47341E+02 &  fixed              \\
$b^{+} (\text{GeV}^{-2})$              &  0.00000E+00 & 0.00000E+00 & 0.00000E+00 & fixed              \\
\hline                                                                                         
$\alpha_{-}(0)$                        &  0.32715E+00 & 0.13892E-01 & 0.32715E+00 & fixed              \\
$\alpha_{-}^{'} (\text{GeV}^{-2})$  &  0.11000E+01 & 0.33881E-01 & 0.11000E+01 & fixed              \\
$C^{-}$   (mb)                         &  0.33528E+02 & 0.13387E+01 & 0.33528E+02 &  fixed              \\
$b^{-} (\text{GeV}^{-2})$               &  0.00000E+00 & 0.00000E+00 & 0.00000E+00 &  fixed             \\
\hline                                                                                         
$H_{1}$  (mb)                                 &  0.31370E+00 & 0.16934E-03 & 0.33974E+00 &  0.14696E-03 \\
$H_{2}$ (mb)                                  & -0.21950E+01 & 0.12102E-01 & 0.27105E+01 &  0.50719E-02   \\
$H_{3}$  (mb)                                 &  0.39935E+02 & 0.98913E-01 & 0.50953E+02 &  0.62230E-01\\
$b^{H}_{1} (\text{GeV}^{-1})$               &  0.25927E+01 & 0.97184E-03 & 0.26824E+01 &  0.82689E-03\\
$b^{H}_{2} (\text{GeV}^{-1})$               &  0.72045E+01 & 0.27693E-01 & 0.61736E+01 &  0.13102E-01\\
$b^{H}_{3} (\text{GeV}^{-1})$               &  0.48405E+01 & 0.10107E-01 & 0.44076E+01 &  0.52826E-02\\
$r_{+} (\text{GeV}^{-1})$                       &  0.26818E+00 & 0.57931E-04 & 0.26436E+00 &  0.50348E-04\\
\hline                                                                                          
$O_{1}$  (mb)                                  & -0.44278E-01 & 0.20397E-03 & 0.42841E-01 &  0.17151E-03\\
$O_{2}$   (mb)                                 &  0.93254E+00 & 0.14218E-01 & 0.83063E+00 &  0.14265E-01\\
$O_{3}$   (mb)                                 & -0.17655E+02 & 0.80820E-01 & 0.17510E+02 &  0.76993E-01\\
$b^{O}_{1} (\text{GeV}^{-1})$                &  0.15832E+01 & 0.41271E-02 & 0.15684E+01 &  0.38186E-02\\
$b^{O}_{2} (\text{GeV}^{-1})$                &  0.28034E+01 & 0.20216E-01 & 0.26724E+01 &  0.19453E-01\\
$b^{O}_{3} (\text{GeV}^{-1})$                &  0.28929E+01 & 0.59137E-02 & 0.28842E+01 &  0.56380E-02\\
$r_{-} (\text{GeV}^{-1})$                        &  0.26818E+00 & 0.57931E-04 & 0.26436E+00 &  0.50348E-04\\    
                                                                                   
	\end{tabular}                                                                     
	\caption{Parameters of standard Pomeron and Odderon, of their double rescatterings, of secondary Reggeons  and their errors in FMO  model determined from the fits to the data on $d\sigma/dt$. Total cross sections $\sigma_{tot}$ and ratios $\rho$  were included in the fit without the Coulomb term}
\end{table*}



\section{Slope $B(s, t)$} \label{sec:10}
The slope $B(s, t)$ is a very interesting quantity in the search for Odderon effects.
It is defined by
\begin{equation}\label{eq:slope(s,t)}
B(s,t)=\dfrac{d}{dt}\ln(d\sigma/dt).
\end{equation}

\begin{figure*}
	\includegraphics{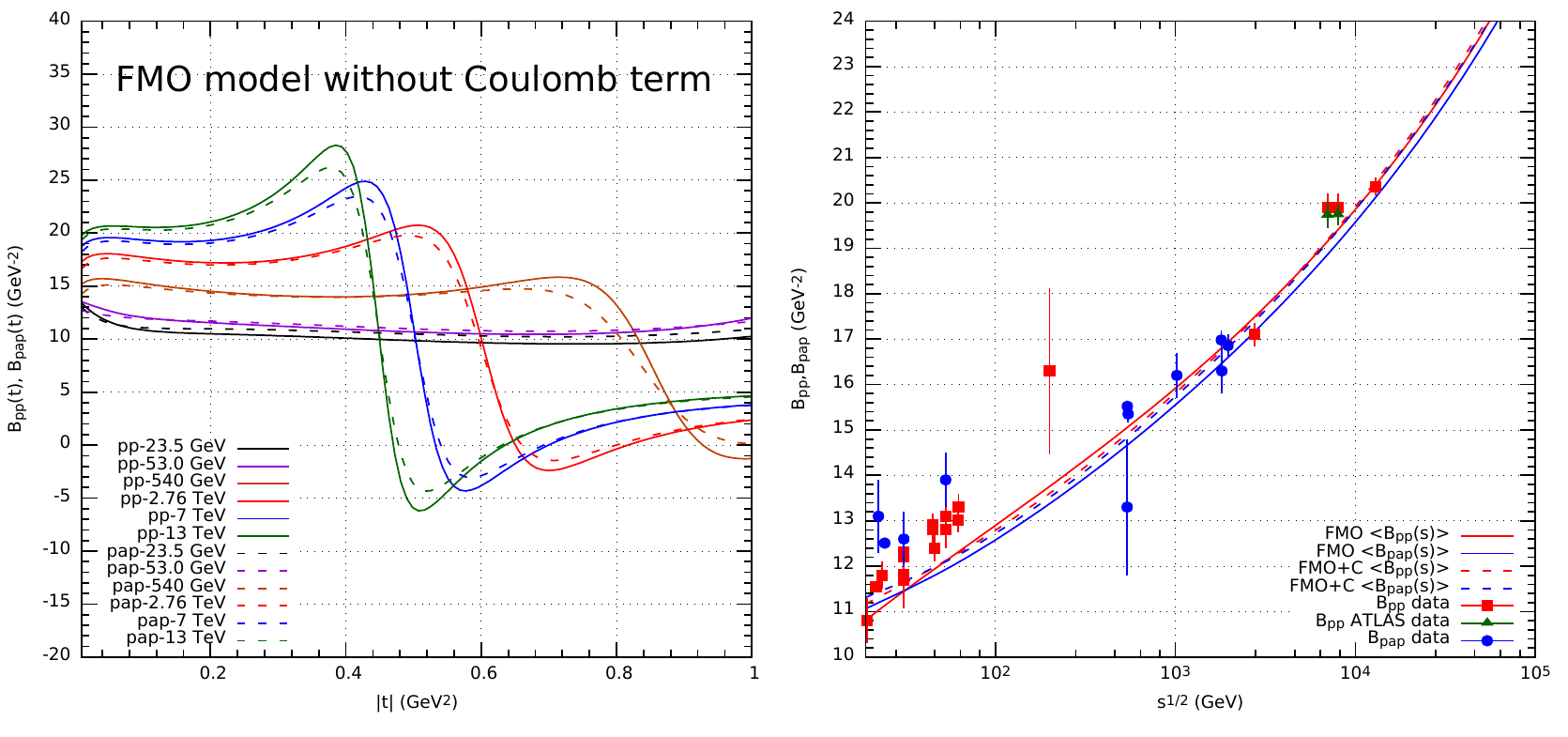}
	\caption{Slopes $B^{pp}(t)$ and $B^{\bar pp}(t)$ at increasing energy(left panel) and the $s$-dependence of the averaged slopes $<B^{pp}(s)>$,  $<B^{\bar pp}(s)>$  together with experimental data (right panel)}
	\label{fig:B(s)B(t)multi}       
\end{figure*}

If we consider the dependence of slope on energy and compare this dependence with available experimental data we have to take into account that slopes in any realistic model depend on $t$. Dependence of slope on $t$ at various energies in the FMO model is illustrated in Fig.  \ref{fig:B(s)B(t)multi} (left panel).  Therefore we must to calculate the slope $<B(s)>$ averaged in some interval of $t$. We did that in the interval $|t|\in (0.05, 0.2) \text{GeV}^2$ for GeV energies which approximately is in agreement to the intervals from which the experimental data on $B$ are determined. 
\begin{equation}\label{eq:averagedB}
\begin{array}{ll}
<B(s)>&=(1/\Delta_t)\int \limits _{t_{max}}^{t_{min}}dt \,\dfrac{d}{dt}(\ln(d\sigma(t)/dt))\\
&= (1/\Delta_t )\left  [\ln\left (\dfrac{d\sigma(t_{min})/dt}{d\sigma(t_{max})/dt}\right )\right ]
\end{array}
\end{equation}
where $\Delta_t=t_{max}-t_{min}$.


We show in Table 4 our predictions for the averaged slopes in the TeV region of energy as compared  with experiments at Tevatron and LHC.

\begin{table*} \label{tab:7}
\centering
\begin{tabular}{llcccc}
	\hline
	Energy (TeV)           &Experiment & \multicolumn{2}{c} {$<B^{pp}(s)>$ (GeV$^{-2}$)} & \multicolumn{2}{c}{ $<B^{\bar pp}(s)>$  (GeV$^{-2}$)} \\ \hline
	                       & & Experimental data &          FMO model          & Experimental data &             FMO model             \\
1.8    & E710&         -         &                 16.70            &   16.3$\pm$0.5    &               16.39              \\
1.8     &CDF &         -         &                 16.70            &  16.98$\pm$0.25   &            16.39      \\
1.96   & D0 &         -         &                   16.84            &  16.86$\pm$0.25   &               16.537              \\
2.76  &TOTEM &   17.1$\pm$0.26   &          17.43            &        -       &            17.13               \\
7   &  TOTEM &   19.9$\pm$0.3    &            19.18           &         -         &               18.91               \\
7   & ATLAS  &  19.73$\pm$0.39   &            19.18            &         -         &               18.91               \\
8    &TOTEM  &   19.9$\pm$0.3    &            19.45            &         -         &                19.19               \\
8     & ATLAS&  19.74$\pm$0.31   &            19.45            &         -         &                19.19               \\
13  &  TOTEM &  20.4$\pm0.01$   &           20.50            &            -           &         20.25               \\
\end{tabular}
\caption{Experimetal values of slopes of $pp$ and $\bar pp$ differential cross sections at TeV energies and the averaged slopes calculated in FMO model} 
\end{table*}	

In Fig. \ref{fig:B(s)B(t)multi} (right panel) we show  the increasing of the averaged slopes at t=0 with increasing energy. One can see that the slopes are approaching the $\ln^2s$ increase at high energies. 


In Fig. \ref{fig:B(s_exp,t)} we plot the slopes as function of $t$ in $pp$ and $\bar pp$ scatterings. We discover from the $t$-dependence of the slopes an extremely interesting phenomenon. The slope in $pp$ scattering has a  different behaviour in $t$ than the slope in  $\bar pp$ scattering. 
In the left panel of Fig. \ref{fig:B(s_exp,t)}  we see that in $pp$ scattering  the slopes are  first nearly constant and  after that they fall sharply, they cut a first time the $B(t)=0$ line, reach a deep minimum  negative value, after that they increase and cut a second time the $B(t)=0$  line and finally they reach an approximately constant value for higher $t$. The two crossing points of the  $B(t)=0$ line move towards smaller $t$ when energy increases.
In the right panel of Fig. \ref{fig:B(s_exp,t)} we see a very different behaviour in $\bar pp$ scattering.  In this case, at energies higher than ISR ones, B(t) marginally crosses zero, but no so deeply and sharply as in pp scattering.
For completeness, we show in Fig. \ref{fig:B(s,t)7-13pp,1_96pap}  the slope parameter  for $pp$ scattering at 7 and 13 TeV as compared with  the slope parameter in $\bar pp$ scattering at 1.96 TeV, where we can see the same phenomenon.

This phenomenon is a clear Odderon effect.  The odd-under crossing amplitude  makes the difference between $pp$ and $\bar pp$ scatterings and this amplitude is dominated at high energy by the Maximal Odderon.

\section{Comparison with other approaches}
To our knowledge, the present model is the only model which fits forward and non forward  data in a wide range of energies (including TeV region), without theoretical defects (like the violation of the unitarity).

However, it is important to note that our results concerning the slopes are in complete agreement with those obtained recently by Cs\"{o}rg\"{o} {\it et al.}  \cite{Csorgo}, who performed a very useful mirroring between the discontinuous experimental data (points) and continuous analytic functions (scattering amplitudes) by using an expansion in terms of L\'{e}vy polynomials. In such a way they get a very clear Odderon effect concerning the slopes. Their analysis have no dynamical content: it is a parametrization of experimental data in terms of big number of  parameters. 

This agreement is very important from two points of view. On one side, the Odderon existence is reinforced by two quite different analysis, one model-independent and the other one having a dynamical content.

On another side, the fact that the Maximal Odderon is in agreement with a model-independent analysis reinforce the status of the Maximal Odderon.

\section{Conclusion}
\label{sec:11}
In our paper we present an extension of the Froissaron-Maximal Odderon (FMO) approach for $t$ different from zero, which satisfies rigorous theoretical constraints.
Our extended FMO approach gives an excellent description of the 3266  \footnote{Experimental data at $t=0$ were taken from \cite{PDG}, with the recent TOTEM and ATLAS points being added.  Set of data at $t\neq 0$ will be send after personal request to E. Martynov.} experimental points considered in a wide range of energies and momentum transferred.
One spectacular theoretical result is the fact that the difference in the dip-bump region between $\bar pp$  and  $pp$ differential cross sections  is diminishing with increasing energies and for very high energies (say 100 TeV), the difference in the dip-bump region between $\bar pp$  and  $pp$ 
is changing its sign: $pp$  becomes bigger than $\bar pp$ at $|t|$ about 1 GeV$^2$. This is a typical Odderon effect.

Another important - phenomenological - result of our approach  is that the slope in $pp$ scattering has a different behaviour in $t$ than the slope in  $\bar pp$  scattering.  This is a clear Odderon effect.

Let us emphasize  that the FMO model is in a good agreement with the data in a wide interval  of energy. However, there is a some discrepancy of the data and model in a region around $\sqrt{s}$=2 TeV  (it is illustrated in the  Fig. \ref{fig:pp-2.76-pap-1.8-FMO}). At the same time agreement with the data at lower and at higher energies is really very good. This problem requires a special investigation
which we will perform after the publication of the common TOTEM/D0 paper \cite{TOTEM/D0}.

New ways of detecting Odderon effects, e. g. in an Electron-Ion Collider, were recently explored on the basis of a general QCD light front formalism \cite{Dumitru-et-al}.

{\bf Acknowledgment.} The authors thank Prof. Simone Giani for a careful reading of the manuscript. One of us  (E.M.) thanks the Department of Nuclear Physics and Power Engineering of the National Academy of Sciences of Ukraine for support (continuation of the project No 0118U005343).

\appendix
\section{Appendix }\label{sec:appendicsA}

\subsection{General constraints}\label{App:Gen.Con}
Let us reiterate here that the model with $\sigma_{t}(s)\propto
\ln^{2}s$ is not compatible with a linear pomeron trajectory having
the intercept 1. Indeed, let us assume that
\begin{equation}\label{eq:trajectory}
\alpha_{P}(t)=1+\alpha_{P}'t
\end{equation}
and the partial wave amplitude has the form
\begin{equation}\label{eq:j-gen}
\varphi (j,t)=\eta (j)\frac{\beta (j,t)}{\left
	[j-1-\alpha_{P}'t\right]^{n}}\approx \frac{i\beta (1,t)}{\left
	[j-1-\alpha_{P}'t\right]^{n}},
\end{equation}
\begin{equation}\label{eq:signature-j}
\eta (j)=\frac{1+\xi e^{-i\pi
		j}}{-\sin\pi j}.
\end{equation}

For Pomeron (simple or double pole) and Froissaron signature is positive, $\xi=+1$.

In ($s,t$)-representation amplitude $\varphi (j,t)$ is transformed to
\begin{equation}\label{eq:s,t-gen}
a(s,t)=\frac{1}{2\pi i}\int dj \varphi (j,t)e^{\xi j}, \quad
\xi=\ln(s/s_{0}).
\end{equation}
Then, we have pomeron contribution at large $s$ as
\begin{equation}\label{eq:s,t-gen1}
a(s,t)\approx -\tilde \beta(t)[\ln(-is/s_{0})]^{n-1}(-is/s_{0})^{1+\alpha'_{P}t}
\end{equation}
where
\begin{equation}\label{eq:tide beta}
\tilde \beta(t)=\beta (t)/\sin(\pi \alpha_{P}(t)/2).
\end{equation}
If as usually  $\tilde \beta(t)=\tilde \beta\exp(bt)$ then we obtain
\begin{eqnarray}\label{eq:sigma-0}
\sigma_{t}(s)&&\propto \ln^{n-1}s,\nonumber\\
\sigma_{el}(s)&&\propto
\frac{1}{s^{2}}\int \limits_{-\infty}^{0}dt|a(s,t)|^{2}\propto
\ln^{2n-3}s.
\end{eqnarray}
According to the obvious inequality,
\begin{equation}\label{eq:unitbound}
\sigma_{el}(s)\leq \sigma_{t}(s)
\end{equation}
we have
\begin{equation}\label{eq:nbound}
2n-3\leq n-1 \qquad  \Rightarrow  \qquad n\leq 2.
\end{equation}

Thus we come to the conclusion that the {\it  a model with $\sigma_{t}(s)\propto
	\ln^{2}s$  (n=3) is incompatible with a linear po\-me\-ron trajectory}. In other words  the partial amplitude Eq. (\ref{eq:j-gen}) with $n=3$ is incorrect.

If $n=1$ we have a simple $j$-pole leading to constant total cross section and vanishing at $s\to \infty$ elastic cross section. However such a behaviour of the cross sections is not supported by experimental data.

If $n=2$ we have the model of dipole pomeron ($\sigma_{t}(s)\propto
\ln(s)$) and would like to emphasize that double $j$-pole is the maximal singularity of partial amplitude settled by unitarity bound
(\ref{eq:unitbound}) if its trajectory is linear at $t\approx 0$. 

We would like to notice here that {\it TOTEM data for the $pp$ total cross section exclude the dipole pomeron model} which is unable to describe with a reasonable $\chi^2$ the high values of $\sigma^{pp}_{tot}(s)$ at LHC energies.

Thus, constructing the model leading to cross section which increases faster than $\ln(s)$, we need to consider a more
complicated case (we consider at the moment a region of small $t$ and $j\approx 1$):
\begin{equation}\label{eq:j-gen**2}
\begin{array}{ll}
\varphi_+(j,t)&=\dfrac{\beta (j,t)}{\left
	[j-1+r(-t)^{1/\mu }\right]^{n}}\\ 
&\approx \dfrac{i\beta (1,t)}{\left
	[j-1+r(-t)^{1/\mu }\right]^{n}}.
\end{array}
\end{equation}
Making use of the same arguments as above, we obtain
\begin{equation}\label{eq:nmbound}
\sigma_{t}(s)\propto \ln^{n-1}s,
\end{equation}
\begin{equation}\label{eq:nmbound-2}
\sigma_{el}(s)\propto
\ln^{2n-2-\mu }s \qquad \mbox{and} \qquad \mu \geq n-1. 
\end{equation}
However in this case amplitude $a(s,t)$ has a branch point at $t=0$ which is forbidden  by analyticity of amplitude $a(s,t)$.

A proper form of amplitude leading to $t_{eff}$ 
\footnote{$t_{eff}$ can be defined by behaviour of elastic scattering amplitude at $s\to \infty $. If  $a(s,t)\approx sf(s)F(t/t_{eff}(s))$ then $\sigma_{el}(s)\propto|f(s)|^{2}\int_{-\infty}^{0}dt|F(t/t_{eff})|^{2}=
	t_{eff}|f(s)F(1)|^{2}$.} 
decreasing faster than $\ln^{-1}s$ (it is necessary for $\sigma_{t}$ rising faster than $\ln s$) is the following
\begin{equation}\label{eq:j-gencorrect}
\varphi_+ (j,t)=\dfrac{\beta (j,t)}{\left
	[(j-1)^{m}-rt\right]^{n}}.
\end{equation}
Now we have $m$ branch points colliding at $t=0$ in $j$-plane and creating the pole of order $mn$ at $j=1$ (but there is no branch point in $t$ at $t=0$). At the same time $t_{eff}\propto 1/\ln^{m}s$ and from $\sigma_{el}\propto \ln^{2mn-2-m}s\leq \sigma_{t}\propto \ln^{mn-1}s\leq
\ln^{2}s$ one  obtains
\begin{equation}\label{eq:mnbound}
\left \{
\begin{array}{ll}
mn &\leq m+1, \\
mn &\leq 3.
\end{array}
\right .
\end{equation}
If $\sigma_{el}\propto \sigma_{t}$ then $n=1+1/m$. Furthermore, if $\sigma_{t}\propto \ln s$ then $m=1$ and $n=2$ which corresponds just to the dipole pomeron model. In the Froissaron (or tripole pomeron) model $m=2$ and $n=3/2$. It means that $\sigma_{t}\propto \ln^{2}s$.

\subsection{Partial amplitudes}\label{App:partampl}
As it follows from Eq.(\ref{eq:mnbound}) for the dominating at $s\to \infty $
contribution in a Froissaron model with $\sigma_{t}(s)\propto
\ln^{2}(s)$, i.e. $n=2$, $m=3/2$, we have to take (here and in what follows we used a more convenient notations $\omega=j-1$ and $\omega_{0\pm }=r_\pm\tau=r_\pm\sqrt{-t/t_0},\quad t_0=1 \text{GeV}^2$). Then

\begin{equation}\label{eq:tripole1}
\begin{array}{ll}
\varphi_{\pm}(\omega,t)&= \eta_{\pm} (\omega)\dfrac{\beta_{\pm}(\omega,t)}{(\omega^{2}+\omega_{0\pm}^2)^{3/2}}\\ 
&=\binom{i}{1}e^{-i\pi \omega/2} \dfrac{\tilde \beta_\pm(\omega,t)}{(\omega^{2}+\omega_{0\pm}^2)^{3/2}}
\end{array}
\end{equation}
where
\begin{equation}\label{eq:signature factor}
\eta_\pm (\omega)=\frac{1\mp e^{-i\pi\omega} }{\sin\pi\omega}.
\end{equation}
For even signature
\begin{equation}\label{eq:beta+}
\tilde \beta_+(\omega,t)=\beta_+(\omega,t)/\cos(\omega \pi/2)
\end{equation}
and for odd signature 
\begin{equation}\label{eq:beta-}
\tilde \beta_-(\omega,t)=\beta_-(\omega,t)/\sin(\omega \pi/2).
\end{equation}

Now let us suppose that in agreement with the structure of the singularity of $\phi_\pm(\omega,t)$ at $\omega^2+\omega_{0\pm}^2=0$ the functions $\tilde \beta_\pm(\omega,t)$ depend on $\omega$ through the variable $\kappa_\pm =(\omega^2+\omega_{0\pm}^2) ^{1/2}$  and it can be expanded in powers of $\kappa_\pm$
\begin{equation}\label{eq:phi-pm}
\phi_\pm(\omega,t)=\binom{i}{1}e^{-i\pi \omega/2}\dfrac{\tilde \beta_{1\pm}(t)+\kappa_\pm \tilde \beta_{2\pm}(t)+ \kappa_\pm^2 \tilde \beta_{3\pm}(t)}{\kappa_\pm ^{3}}.
\end{equation}

There are a different ways to add to partial amplitude $\varphi (j,t)$ terms which at $s\to \infty$ are small  corrections (they can be named as subasymptotic terms). 

Thus we can expand the ``residue'' $\beta(\omega,t) $ in powers of $\omega$ (if $\beta (\omega,t) $ has not branch point in $\omega$ at $\omega=0$) or in powers of $(\omega^2+\omega_0^2)^{1/2}$. Then, for the  first case
\begin{equation}\label{eq:expand1}
\tilde \beta(\omega,t)=\tilde \beta_1(t)+\omega \tilde \beta_2(t)+\omega^2\tilde \beta_3(t),
\end{equation}
and in the second case we have (just this case is explored in the Section \ref{sec:5}) 
\begin{equation}\label{eq:expand2}
\tilde \beta(\omega,t)=\tilde \beta_1(t)+(\omega^2+\omega_0^2)^{1/2} \tilde \beta_2(t)+(\omega^2+\omega_0^2) \tilde \beta_3(t).
\end{equation}

Let us notice that the main terms in $\varphi(j,t)\equiv \varphi(\omega,t)$ for both cases are coinciding having  a pair of branch points colliding at $\omega_0=0\quad  (t=0)$ and generating a triple pole in partial amplitude.  

Taking into account the table integrals
\begin{equation}\label{eq:besselgen}
\int \limits_{0}^{\infty} dx x^{\alpha-1}e^{-\omega
	x}\textit{J}_{\nu }(\omega_{0}x)=I_{\nu }^{\alpha }(\omega, \omega_0)
\end{equation}\label{eq:besselgen-1}
where
\begin{equation}
\begin{array}{ll}
I_{\nu }^{\nu+1}&=\dfrac{(2\omega_{0})^{\nu }}{\sqrt{\pi }
	\dfrac{\Gamma (\nu+1/2)}
	{(\omega^{2}+\omega_{0}^{2})^{\nu+1/2}}},\\ \\ 
I_{\nu }^{\nu+2}&=2\omega \dfrac{(2\omega_{0})^{\nu }}{\sqrt{\pi }
	\dfrac{\Gamma (\nu+3/2)}
	{(\omega^{2}+\omega_{0}^{2})^{\nu+3/2}}},
\end{array}
\end{equation}
one can find
\begin{equation}\label{eq:phi1}
\begin{array}{rl}
\dfrac{1}{(\omega^{2}+\omega_{0}^{2})^{3/2}}&=\dfrac{1}{\omega_{0}}
\int\limits_{0}^{\infty}
dx xe^{-x\omega }J_{1}(\omega_{0}x), \\\\
\int\limits_C \dfrac{d\omega}{2\pi i} \dfrac{e^{\xi\omega}}{(\omega^2+\omega_0^2)^{3/2}}& = \dfrac{J_1(\omega_0\xi)}{\omega_0\xi}. 
\end{array}
\end{equation}
\begin{equation}\label{eq:phi2}
\begin{array}{rl}
\dfrac{1}{\omega^{2}+\omega_{0}^{2}}&=\dfrac{1}{\omega_{0}}\int\limits_{0}^{\infty}
dx e^{-x\omega }\sin(x\omega_{0}),  \\\\
  \int\limits_C \dfrac{d\omega}{2\pi i} \dfrac{e^{\xi\omega}}{\omega^2+\omega_0^2} &= \dfrac{\sin(\omega_0\xi)}{\omega_0\xi}. 
\end{array}
\end{equation}
\begin{equation}\label{eq:phi3}
\begin{array}{rl}
\dfrac{1}{(\omega^{2}+\omega_{0}^{2})^{1/2}}&=\int\limits_{0}^{\infty}
dx e^{-x\omega }J_{0}(\omega_{0}x), \\ \\
\int\limits_C \dfrac{d\omega}{2\pi i} \dfrac{e^{\xi\omega}}{(\omega^2+\omega_0^2)^{1/2}}& = J_0(\omega_0\xi). 
\end{array}
\end{equation}

\end{document}